%% file: sample-acmsmall.tex
\DocumentMetadata{}
\documentclass[manuscript, screen]{acmart}
\AtBeginDocument{%
  }

\setcopyright{acmlicensed}
\copyrightyear{2025}
\acmYear{2025}
\acmDOI{XXXXXXX.XXXXXXX}

\usepackage{diagbox}
\usepackage{footnote}
\makesavenoteenv{tabular}
\makesavenoteenv{table}
\usepackage{xargs} 
\usepackage{soul}  
\usepackage{color} 
\usepackage{multirow} 
\usepackage{xspace}
\usepackage{listings}
\usepackage{makecell}
\usepackage{enumitem}
\usepackage{verbatim}
\usepackage{tabularx}
\usepackage{multirow}
\usepackage{appendix}
\usepackage{subcaption}

\newcommand{\eg}{{\it e.g.,\ }}

\newcommand{\etc}{{\it etc.}}
\newcommand{\ie}{{\it i.e.,\ }}

\usepackage{booktabs}
\usepackage{array}
\definecolor{oxfordblue}{rgb}{0.0, 0.13, 0.28}
\definecolor{harvardcrimson}{rgb}{0.79, 0.0, 0.09}
\definecolor{dartmouthgreen}{rgb}{0.05, 0.5, 0.06}
\definecolor{princetonorange}{rgb}{1.0, 0.56, 0.0}
\definecolor{yaleblue}{rgb}{0.06, 0.3, 0.57}
\definecolor{usccardinal}{rgb}{0.6, 0.0, 0.0}
\definecolor{uclablue}{rgb}{0.33, 0.41, 0.58}
\definecolor{msugreen}{rgb}{0.09, 0.27, 0.23}
\definecolor{cornellred}{rgb}{0.7, 0.11, 0.11}
\definecolor{pomegranate}{RGB}{192, 57, 43}
\definecolor{anti-pomegranate}{RGB}{43,178,192}
\definecolor{alizarin}{RGB}{231, 76, 60}
\definecolor{anti-belize}{RGB}{185, 41, 56}
\definecolor{belize}{RGB}{41, 128, 185}
\definecolor{peter}{RGB}{52, 152, 219}
\definecolor{green}{RGB}{22, 160, 133}
\definecolor{anti-green}{RGB}{160,22,118}
\definecolor{turquoise}{RGB}{26, 188, 156}
\definecolor{pumpkin}{RGB}{211, 84, 0}
\definecolor{anti-pumpkin}{RGB}{0,22,211}
\definecolor{carrot}{RGB}{230, 126, 34}
\definecolor{wisteria}{RGB}{142, 68, 173}
\definecolor{anti-wisteria}{RGB}{99,173,68}
\definecolor{amethyst}{RGB}{155, 89, 182}
\definecolor{nephritis}{RGB}{39, 174, 96}
\definecolor{anti-nephritis}{RGB}{174,39,117}

\newcommand{\penguin}[1]{{\color{black} #1}}
\newcommand{\zhenhui}[1]{{\color{black} #1}}
\newcommand{\pzh}[1]{{\color{black} #1}}
\newcommand{\cxy}[1]{{\color{black} #1}}
\newcommand{\xinyue}[1]{{\color{black} #1}}
\newcommand{\xy}[1]{{\color{black} #1}}
\newcommand{\data}[1]{{\color{black} #1}}
\newcommand{\peng}[1]{{\color{black} #1}}
\newcommand{\name}{{\textit{DesignLearner}}}

\newcommand{\cx}[1]{{\color{black} #1}}
\newcommand{\hwx}[1]{{\color{black} #1}}
\newcommand{\zhj}[1]{{\color{black} #1}}
\newcommand{\re}[1]{{\color{black} #1}}
\newcommand{\mr}[1]{{\color{black} #1}}
\newcommand{\rr}[1]{{\color{black} #1}}
\begin{document}

\title[DesignLearner -- Redesign of Online Design Communities]{Redesign of Online Design Communities: Facilitating Personalized Visual Design Learning with Structured Comments}

\acmConference[CSCW'25]{In Proceedings of The 2025 ACM
SIGCHI Conference on Computer-Supported Cooperative Work & Social Computing (Conference CSCW ’25)}{October 2025}{Bergen, Norway}
\author{Xia Chen}
\authornotemark[1]
\email{chenx776@mail2.sysu.edu.cn}
\affiliation{%
  \institution{Sun Yat-Sen University}
  \city{Zhuhai}
  \country{China}
}

\author{Xinyue Chen}
\authornote{Both authors contributed equally to this research.}
\email{chenxy857@mail2.sysu.edu.cn}
\affiliation{%
  \institution{Sun Yat-Sen University}
  \city{Zhuhai}
  \country{China}
}

\author{Weixian Hu}
\email{huwx25@mail2.sysu.edu.cn}
\affiliation{%
  \institution{Sun Yat-Sen University}
  \city{Zhuhai}
  \country{China}
}

\author{Haojia Zheng}
\email{zhenghj36@mail2.sysu.edu.cn}
\affiliation{%
  \institution{Sun Yat-Sen University}
  \city{Zhuhai}
  \country{China}
}

\author{Yujun Qian}
\email{qianyj8@mail2.sysu.edu.cn}
\affiliation{%
  \institution{Sun Yat-Sen University}
  \city{Zhuhai}
  \country{China}
}

\author{Zhenhui Peng}
\authornote{Corresponding author.}
\email{pengzhh29@mail.sysu.edu.cn}
\affiliation{%
  \institution{Sun Yat-Sen University}
  \city{Zhuhai}
  \country{China}
}

\renewcommand{\shortauthors}{Xia Chen et al.}

\begin{abstract}
Online Design Communities (ODCs) offer various artworks with members' comments for beginners to learn visual design. However, as identified by our Formative Study (N = 10), current ODCs lack features customized for personal learning purposes, \eg searching artworks and digesting useful comments to learn design principles about buttons. In this paper, we present \name{}, a redesigned interface of ODCs to facilitate personalized visual design learning with comments structured based on UI components (\eg button, text) and visual elements (\eg color, contrast). In \name{}, learners can specify the UI components and visual elements that they wish to learn to filter artworks and associated comments. They can interactively read comments on an artwork, take notes, and get suggestions for the next artworks to explore. Our between-subjects study (N = 24) indicates that compared to a traditional ODC interface, \name{} can improve the user learning outcome and is deemed significantly more useful.
We conclude with design considerations for customizing the interface of online communities to satisfy users' learning needs.
\end{abstract}




\maketitle

\input{sections/1_introduction}

\input{sections/2_related_work}

\input{sections/3_formative_study}
\input{sections/4_Design_and_Implementation}

\input{sections/5_Experiment}

\input{sections/6_Analysis_and_Result}

\input{sections/7_Discussion}
\input{sections/8_Conclusion}


\bibliographystyle{ACM-Reference-Format}
\bibliography{references}

\input{sections/0_Appendix}
\end{document}

%% file: sections/1_introduction.tex
\section{Introduction}
People, either majoring in design-related subjects or not, can gain benefits by learning visual design knowledge, \eg about proper usage of color and contrast in buttons \cite{geraci2002designing}. 
For example, the basic color harmony principles can help people develop visually appealing slides in PowerPoint and Keynote for their course projects. 
\re{
This paper focuses on facilitating the learning of design knowledge points about UI components (\eg text, image, button) and visual elements (\eg shape, color, space \cite{scheidt2004common}) by browsing online design communities (ODCs).} 
\zhenhui{For one thing, ODCs like Reddit \rr{\textit{r/UI\_design}} offer a public place with rich examples and diverse peer feedback related to the enactment of these knowledge points on the examples  \cite{gonccalves2014inspires,van2010example,herring2009getting, DQ2023}. 
For another, learning by browsing online design examples is a common practice for novice designers \cite{yen2022seeking,yang2015uncovering,krishna2017increasing,gonccalves2014inspires,van2010example,herring2009getting}.} 
Our primary targeted user groups are the design novices who only take the shared examples and feedback in ODCs as learning materials (merely readers) \cite{raith2019students} but do not directly interact with others via posting and commenting \cite{song2023influences}. 
By exploring the resources in ODCs, novices can develop domain knowledge in visual design, which can be reflected by the applications of the learned knowledge to their visual design practices \cite{goldschmidt2001visual,DQ2023}.

\penguin{
Despite the benefits,
ODCs often lack features that facilitate personalized learning. 
Research on personalized learning commonly aims to accommodate individual learning interests (\eg topics or subjects that learners are curious to explore) and needs (\eg knowledge or skills that learners must acquire to achieve specific goals) \cite{bernacki2021systematic}. 
For example, in our focused informal learning settings in ODCs, learners can have an interest in the UI component ``button'' and need to acquire knowledge about the button's color and shape to perform well in their future design tasks. 
Previous Human-Computer Interaction researchers often implement two approaches to support personalized learning. 
First, they deliver adaptive content that accommodates users' interests, \eg 
based on any user-interested vocabulary list \cite{storyfier_uist23}, typing inputs to generative models \cite{vocabulary_learning_chi24}, and taking photos \cite{draxler2023relevance}. 
Second, they support interactive content management to satisfy users' needs for knowledge acquisition, \eg 
note-taking that helps users consume learning materials \cite{PlanHelper,cao2022videosticker}. 
Existing interfaces of ODCs like those in Reddit \footnote{\eg \url{https://www.reddit.com/r/graphic_design/}, \url{https://www.reddit.com/r/UI_Design/}} and Dribbble \footnote{\url{https://dribbble.com/}} allow users to specify search keywords and view the returned post and comment lists, but they fail to organize this content based on the design knowledge points of learners' interests, \eg about certain UI components and visual elements. 
Additionally, current ODCs
miss the opportunities to facilitate users in managing this content to develop their domain knowledge. 
Our vision is a learner-oriented mode of online design communities where novices can pick up their interested knowledge points by efficiently exploring and managing the visual design examples and comments. 
}

\penguin{
In this paper, we introduce
\name{}, a redesigned interface of online design communities, that facilitates personalized visual design learning with structured comments. 
\name{} is inspired by previous work on online communities and personalized learning. 
For example, \citet{DQ2023} proposed a computational workflow to process comments in Reddit \rr{\textit{r/UI\_design}} and detect the keywords about UI components and visual elements in the comments. 
This workflow can enable us to build a taxonomy of visual design knowledge to enable adaptive content delivery that caters to learners' interests. 
\citet{PlanHelper} developed a tool with a note-taking panel to satisfy users' needs for managing useful community content for constructing their personal plans. 
Despite that previous work has informed potential useful features for our redesigned ODC interface, little is known about what practical challenges users with personalized learning goals would encounter when exploring content in ODCs, how to customize these features to mitigate these challenges, and how the re-designed ODC affect the learning process and outcome.  
Filling these gaps can provide CSCW researchers with insights into supporting personalized learning activities within online communities.
}

To this end, we first conduct a Formative Study with 10 users who are novices in visual design to understand the challenges of learning design knowledge \zhenhui{in Reddit \rr{\textit{r/UI\_design}}, a representative ODC with many comments talking about UI components and visual elements}. 
Users indicated challenges in filtering posts based on knowledge points, reading constructive comments of learning interests efficiently, and managing the learned knowledge systematically.
Then, we build \name{}, a web interface of ODCs with features redesigned for facilitating personalized visual design learning. 
\penguin{Empowered by the workflow in \cite{DQ2023}, we structure each comment in Reddit \rr{\textit{r/UI\_design}} via a taxonomy of visual knowledge, \ie ``UI components - a specific UI component (\eg button) - a specific visual element (\eg color) - a comment
that this visual element and this UI component co-exist''}.
\zhenhui{Based on these structured comments, \name{} embeds an interactive note-taking panel that allows users 1) add any interested comment to as a node with auto-generated knowledge taps into a mindmap and 2) click any node in the mindmap to direct to the source post and comment.
Apart from this unique feature, \name{} supports users to filter posts and comments by choosing the types of UI components and visual elements they wish to learn}. 
When reading the comments, users can view the highlighted knowledge keywords in the comment text.

We conduct a between-subjects study with 24 novices of visual design to evaluate the \name{}'s effectiveness and user experience compared to the \rr{\textit{r/UI\_design}} web page baseline. 
The results show that \name{} improved participants' efficiency in exploring design examples and comments that are helpful for their personalized learning goals. 
Participants with \name{} made significant improvements in designing UIs that reveal the principles of hierarchy and contrast, 
outperforming those who used baseline interface in the visual design task for mobile payment pages.
Participants deem \name{} significantly more useful for personalized visual design learning and favor its features of filtering cases based on knowledge points, structured comments, keyword highlighting of knowledge points, and interactive note-taking on a mindmap.
Based on the results, we discuss the implications of customizing the features of online communities to support personalized learning tasks. 

Our work makes three contributions. 
First, we present \name{}, a redesigned interface of online design communities with features for facilitating personalized visual design learning. 
Second, we extend empirical understandings of how people learn knowledge with user-generated content in online communities. 
Third, we provide insights into future design and applications of online communities to satisfy users' learning goals.

%% file: sections/2_related_work.tex
\section{Related Work}
\subsection{Online Design Communities}

An online design community (ODC, e.g., Reddit 
 \rr{\textit{r/UI\_design}}) refers to a virtual space or platform where individuals including professionals, students, and enthusiasts of design gather to share, discuss, and collaborate on various aspects of design \cite{cheng2020critique}. 
Previous Human-Computer Interaction (HCI) work on ODCs has largely focused on the members who actively participate in the communication in the communities \cite{Jennifer2014, Karen2013, Sayamindu2016}. 
For example, from the perspectives of feedback seekers, \citet{cheng2020critique} studied how creators publicly request feedback in an ODC. 
\citet{Qingyu2023} quantified how the creation stages (\eg work-in-progress or complete) of shared artworks impact the creators' engagement with the received comments. 
As for feedback providers, \citet{Dannels_Gaffney_Martin_2016} examined how the feedback provider's comments reflected his or her communication skills. 
Through exchanging feedback on the artworks, participants of the communication can get feedback for improving their work, connect with others, and gain knowledge of design \cite{yang2015uncovering,krishna2017increasing}.  

\cx{
Our work, unlike previous research, focuses on browsers who utilize ODCs as learning resources for visual design, rather than active participants in the community. 
ODCs, similar to other domains such as creative writing \cite{campbell2016thousands} and programming \cite{informal_learning_chi2022}, facilitate learning through shared design examples and associated feedback. 
However, limited research has addressed how to help users efficiently filter and \penguin{manage the content that matches their interests and needs}. While existing ODC interfaces allow searching and viewing based on user-specified keywords, popularity, and upvotes, the browsers could still face challenges in navigating useful posts and comments that \penguin{talk about UI components and visual elements of their interests, given that the posts and comments are of various quality and are not grouped based on knowledge points} \cite{agichtein2008finding, Jennifer2014, Anbang2012, Anbang2014}. 
\penguin{Our work identifies the challenges of these browsers when conducting personalized learning tasks in ODCs and introduces a learner-oriented ODC interface to facilitate them in these tasks.}
}
\penguin{
\subsection{Personalized Learning Support}
Personalized learning is commonly defined as ``... instruction in which the pace of learning and the instructional approach are optimized for the needs of each learner. Learning objectives, instructional approaches, and instructional content (and its sequencing) may vary based on learner needs.'' \cite[p.9]{king2017reimagining}. 
In other words, personalized learning is learner-centric, and the majority of personalization efforts focus on accommodating learners' interests and needs \cite{bernacki2021systematic}. 
Previous HCI researchers actively implemented intelligent tutoring systems with adaptive content delivery to foster learners' interests. 
For instance, to support language learners to master any target word set they wish, Storyfier \cite{storyfier_uist23} generates short stories as contexts, and RetAssist \cite{retassist_dis24} further generates a series of images picturing any story that covers the word set. 
According to the four-phase model of interest development \cite{hidi2006four}, it is necessary to first trigger and maintain situational interest before students can emerge and 
develop a more stable individual interest in a topic. 
\citet{reber2018personalized} argued
that the personalized materials have a higher probability of increasing situational interest than a standardized, one-size-fits-all approach. 

Prior HCI work also proposed interactive tools, many of which are related to note-taking \cite{deb2018enhancing,cao2022videosticker,8798338,PlanHelper,10.1145/3544548.3581551}, to satisfy users' needs for managing materials in their learning activities. 
For example, iVRNote \cite{8798338} is an interactive VR note-taking interface that records the timing of each stroke in notes to assist students in reviewing lectures and allows them to select portions of their notes to revisit corresponding sections of lectures in a VR environment. 
VideoSticker \cite{cao2022videosticker} supports visual note-taking by extracting expressive content and narratives from videos as `object stickers'.
These tools support the documentation of personally interesting content, which can be viewed as a self-regulated learning practice and may reduce extraneous cognitive load when engaging in the learning materials \cite{moos2009note}.
In line with these previous HCI works, we aim to accommodate users' interests and needs for visual design knowledge in ODCs via features for adaptive content delivery and interactive content management. 
We contribute a learner-oriented ODC interface that supports personalized learning with online community data.
}

\penguin{
\subsection{Interactive Tools for Exploring Online Communities }
HCI researchers have explored two common types of interactive tools to help users explore online communities. 
First, many tools help users filter needed information in the communities \cite{Amplifying-the-music-listening-experience, CoArgue, APP-Breast-Cancer}. 
For instance, CoArgue \cite{CoArgue} facilitates navigating claims in question-answering communities by organizing them based on sentiment and relevance.
\citet{Amplifying-the-music-listening-experience} developed a visual analytic system that enables users to explore song-related comments and empathize with listeners through emotion analysis.
Second, many tools help users digest and manage the content in the communities, \eg by automatically highlighting the content that the users might be interested in \cite{PlanHelper, CQAVis}. 
Interactive note-taking tools further support content management by allowing users to save and annotate useful information \cite{Evaluating-the-effect-of-technology-on-note-taking-and-learning, kang2021metamap, cao2022videosticker, PlanHelper,chi2021notecostruct}. 
\mr{
Additionally, social annotation tools enable collaborative content engagement and knowledge construction by allowing users to comment on and discuss shared content directly within the system \cite{hwang2007study,almahmoud2024enhancing,gao2013case}. For example, \citet{hwang2007study} provided a system with functionalities that include highlighting, underlining, adding textual annotation, and online voice recording. 
Research suggests that discussions supported by social annotation tools are often more focused and rich compared with those in discussion boards \cite{brush2023supporting} and threaded discussion forums \cite{van2006affordance}. 
}

To enable such tools for exploring online communities, researchers have employed a variety of computational techniques for modeling textual comments. Text summarization techniques \cite{keikha2014evaluating,song2017summarizing}, classification models
\cite{zhenhui2021, yangseeker, yang2019channel, sharma2018mental, wambsganss2020conversational}, and regression models \cite{krause2017critique,peng2020exploring} are commonly used to extract, evaluate, and predict comment quality. For design communities, \citet{Qingyu2023} analyzed feedback comments based on features like actionability, justification, and specificity, showing their impact on positive user engagement. \citet{DQ2023} fine-tuned NLP models (\eg T5, RoBERTa) and adopted clustering algorithms to structure the comments in UI design communities based on the types of sentences (\ie critique, suggestion, rationale) and keywords (\eg words about UI components and visual elements). 

Inspired by these efforts, we build upon techniques like 
the workflow of \citet{DQ2023}
to structure comments, enabling learners to filter posts and comments based on the UI components and visual elements they are interested in. 
Different from previous tools, our \name{}
specifically targets users who explore online communities for learning purposes. 
While \name{} focuses on supporting individuals in note-taking, it can be extended as a social annotation tool that engages a group of learners, which we discuss in \autoref{sec:discussion_7.1}. 
}

%% file: sections/3_formative_study.tex
\section{Formative Study}

We conducted a formative study with ten novices to understand their challenges and needs for support when browsing design examples and comments in ODCs.  
The results of this study informed the design decisions we made for \name{}.

\cxy{
\subsection{Research Site: Reddit \textit{r/UI\_design} }
\label{Reddit_introduce}

\penguin{
To situate novices in the visual design learning tasks that 
this paper focuses on in ODCs, we follow two key criteria to filter the appropriate research sites. 
First, the ODCs should contain rich shared design examples about web or mobile interfaces 
with various UI components relevant to this paper's focus.
Second, the ODCs should provide constructive feedback on the shared design examples. 
For the popular ODC Dribbble \footnote{\url{https://dribbble.com/}}, we observe that most comments are short and only express appreciation, \eg ``Super work'', which could violate the second criterion. 
For other popular ODCs like Graphic Design Forum \footnote{\url{https://www.graphicdesignforum.com/}} and those subreddits related to visual design \footnote{We use the keyword ``design'' to search communities on Reddit and manually examine the returned top 30 communities to check if they mainly talk about visual design. For example, \rr{\textit{r/InteriorDesign}} is more about the design within a building rather than the web or mobile interfaces.} on Reddit (\eg \rr{\textit{r/design\_critiques}}, \rr{\textit{r/graphic\_design}}, \rr{\textit{r/Design}}, \rr{\textit{r/UXDesign}},
\rr{\textit{r/web\_design}}, and \rr{\textit{r/UXDesign}}), we observe that they either cover a wide range of design topics like branding and visual arts or lack tags that specify the post's purpose for requesting feedback, which makes it challenging to identify the design examples that satisfy our first criterion. 
One exception is Reddit \rr{\textit{r/UI\_design}} \footnote{\url{https://www.reddit.com/r/UI_Design/}}. 
It is a representative ODC with over 179K members sharing public examples and comments about user interfaces for the web and devices as of June 2024, ranking top 1\% of all communities on Reddit in terms of size. 
It contains flairs like ``Feedback Request'' that help us filter relevant design examples and constructive feedback. 
Therefore,
we adopted the Reddit \rr{\textit{r/UI\_design}} webpage as the ODC in our formative study, used as data source of our taxonomy, as well as our baseline ODC interface in the later user study.
}
To minimize distractions from other types of posts, such as ``UI/UX Design Trend Question'', we pre-filtered posts with a ``Feedback Request'' flair for participants before both studies.

}
\subsection{Participants}
\label{sec:for-study-participants}
We recruited 10 students (4 males, 6 females; age range: 19-22, Mean = 20.5, SD = 0.94; labeled as S1-S10) from a local \zhenhui{engineering college} via a post on social media platform. 
The inclusion criteria are that participants have no or little experience in learning visual design ($M = 1.1, SD = 3.73$; 1 - No experience at all, 5 - A lot of experience). 
Eight of them are undergraduates, and two of them are master's students.
They all speak English as their second language and have passed the national College English Test as a general requirement. 
Eight participants are majoring in Artificial Intelligence (AI), one in Software Engineering, and one in Chemical Engineering. 
\zhenhui{We did not recruit students majoring in design-related subjects as they normally have a lot of experience in visual design. 
While the AI or engineering background of our participants can be helpful for refining potential AI features of \name{}, user groups of other majors may contribute more viewpoints for \name{}, which we will discuss in the Limitation subsection.}

\subsection{Task and Procedure}

We adopted the formative study design in \cite{jiang2023graphologue, critrainer, CReBot} to organize our task and procedure.
\penguin{
We first introduced the interface of Reddit \rr{\textit{r/UI\_design}} \cxy{(\autoref{Reddit_introduce})} and the task: 
\textit{``Suppose you are a novice UI designer and you would like to browse the posts and comments in this ODC for twenty minutes to learn about buttons, such as their layout, shape, color, etc. After that, you need to criticize the design of buttons in a UI example''}.
After the task, 
we conducted a semi-structured interview with each participant to reflect on their experiences with guiding questions: 
\textit{``Did you find the comments and posts in this ODC helpful for your learning task?''},
\textit{``Could you describe how you searched for posts and comments and learned from them in your learning task?''}, and 
\textit{``What challenges did you encounter when you explored this ODC in your learning task?''}.
}
We then presented two low-fidelity sketches of a potentially redesigned ODC interface (Figure \ref{fig:formative_study}). 
\re{Five authors of this paper brainstormed the potential features of the redesigned interface based on related work (\eg \cite{DQ2023}, \cite{PlanHelper}), existing ODC interfaces (\eg Dribbble), and their experience in exploring ODCs for learning visual design knowledge. 
For example, the grid layout in the overview page (Figure \ref{fig:formative_study}(\rr{A2})), the layout for displaying a design example and comments, and the recommendation panel (Figure \ref{fig:formative_study}(\rr{B1})) are adapted from the Dribbble interface. 
The categories of UI components and comments are drawn from \cite{DQ2023}, and the panel for note-taking is inspired by \cite{PlanHelper}. }
\penguin{We introduced the potential features one by one and asked the participants,  
\textit{``Please rate your perceived usefulness of our brainstormed features for your learning task on a five-point Likert scale (1 - not useful at all, 5 - very useful) and briefly give your rationale''}. 
Finally, we asked the participants, 
\textit{``Do you have suggestions upon our proposed features, and what other features do you want in the redesigned ODC for personalized visual design learning?''}.
}
Each participant spent about 40 minutes in the whole procedure and got around 5.6 USD in compensation. 
The interviews were recorded, transcribed, and analyzed using the reflexive thematic analysis method \cite{braun2012thematic}. 

\subsection{Findings}
\subsubsection{Challenges of Learning Visual Design via Browsing ODCs}
All participants agreed that the ODC contains useful design examples and associated comments, which help them learn the good and bad design practices of designing buttons in a graphic UI. 
We presented the key categories of the \textbf{C}hallenges (\textbf{C1}-\textbf{C4}) that emerged from the interviews as follows. 

\textbf{Filter Posts based on Knowledge Points.} 
Six participants reported that they struggled to find posts that contain comments about the UI components and visual elements of their learning interests (\textbf{C1}). 
\textit{``I input `button' in the search box and decided whether to read the returned posts in detail based on their titles. However, many comments in my explored posts were not talking about buttons''} (S2, female, age: 20).
Four participants further commented on their difficulties in finding relevant posts that contain comments about similar UI components or visual elements to those in the currently explored post (\textbf{C2}). 
\textit{``When reading the comments of a post, I learned that the button colors should vary based on a button's function. I wanted to check another post to learn more about the button's colors but failed to find it quickly''} (S4, female, age: 19).

\textbf{Read Constructive Comments of Learning Interests Efficiently.}
Five participants found it frustrating to encounter many non-constructive comments in the posts they read in detail (\textbf{C3}). 
\textit{``Several comments in my explored post were constructive critiques on the buttons of the design examples, but most of the comments were unrelated and disturbing''} (S3, male, age: 20). 
\cxy{\textit{``I've noticed that some posts have only one or two lines of useful information in the comments section, and most of the comments are pointless''} (S7, male, age: 22). }

\textbf{Manage the Learned Knowledge Systematically.}
Three participants raised concerns that in the Reddit ODC interface, they were unable to organize and review the learned knowledge in the explored design examples and comments (\textbf{C4}).
\textit{``Many comments are indeed helpful as they talk about what are bad practices in the design example. I want to take notes on these comments in time and to form a knowledge graph. 
However, the interface does not have any related convenient feature for this purpose''} (S7, male, age: 22).  
If participants wanted to record what they had learned in the Reddit ODC, they had to manually copy the image of the design example and textual comments and paste them into the document separate from the current interface, which is time-consuming and \cxy{\textit{``increase the information processing workload''} (S9, female, age: 20).} 

\subsubsection{Perceived Usefulness of Potential Features}

\begin{figure*}[!ht]
    \centering
    \includegraphics[width=1.0\linewidth]{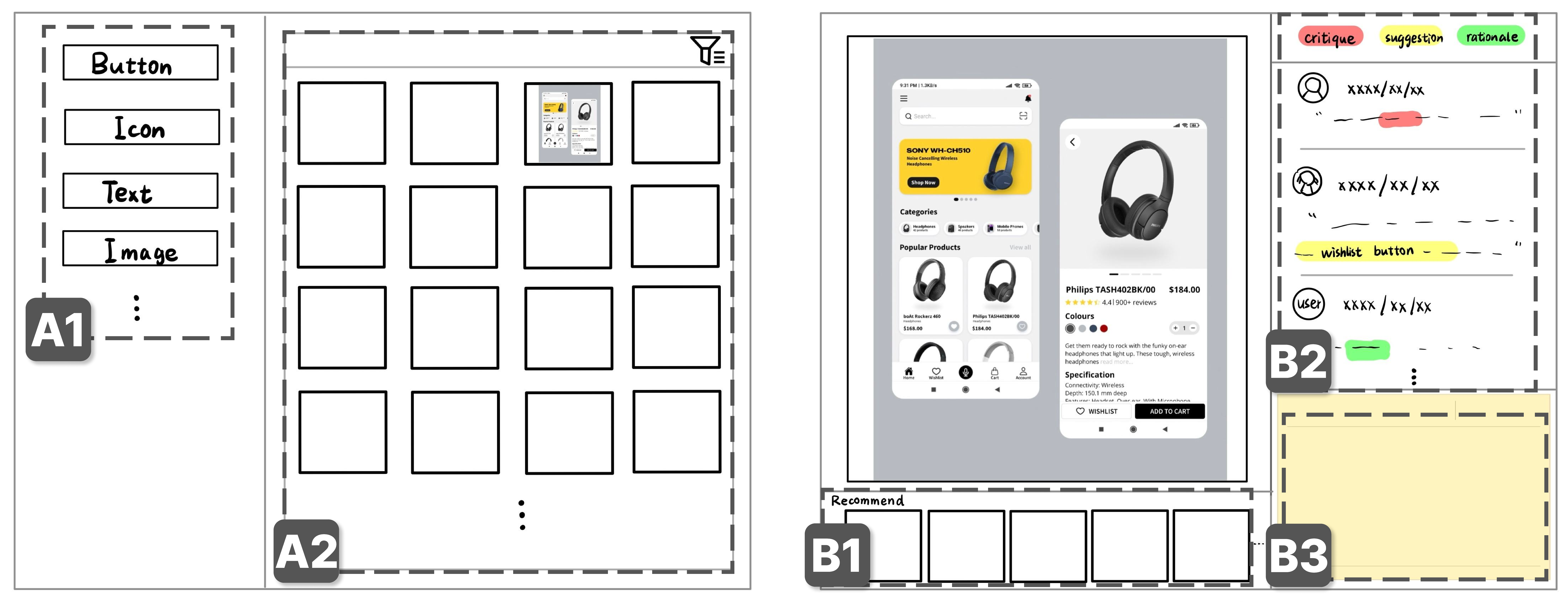}
    \caption{\peng{Features of two low-fidelity sketches of a potential redesigned online design community interface and their perceived usefulness (1 - not useful at all, 5 - very useful) by participants in the formative study. 
    Overview page (left): \rr{A1}) Sort posts based on the UI components that their comments mention ($M = 4.20, SD = 0.62$); 
    \rr{A2}) Further sort posts based on a) the visual elements that their comments mention ($M = 3.90, SD = 0.54$) or b) the creating time of the posts ($M = 3.30, SD = 0.90$). 
    Reading page (right): \rr{B1}) Recommend related posts based on a) the targeted UI component ($M = 4.90, SD = 0.10$) or b) the visual element ($M = 4.10, SD = 0.77$) explored in the current post; 
    \rr{B2}) Highlight a) sentences that provide meaningful feedback ($M = 4.30, SD = 0.68$) or b) keywords of visual elements ($M = 4.40, SD = 0.49$) in the comments; 
    \rr{B3}) Note-taking panel ($M = 4.10, SD = 0.99$).
    }}
    \label{fig:formative_study}
\end{figure*}

\autoref{fig:formative_study} presents the potential features in a potential ODC interface redesigned for learning purposes and the participants' average ratings on their usefulness. 
In general, participants would find it useful if the interface allows them to sort posts based on the UI components (\rr{A1}) and visual elements (\rr{A2}) that the comments of the posts mention. 
On the reading page, it would be useful to recommend related posts to the current one based on the knowledge points of participants' learning interests (\rr{B1}), 
highlight the sentences and keywords that provide meaningful feedback from the aspects of learners' interests (\rr{B2}), and offer a panel for note-taking (\rr{B3}). 
Three participants expected that the note-taking panel should support interaction with the posts and comments, such as \textit{``shortcuts for adding related comments to the panel''} (S1, male, age: 21) and \textit{``buttons for revisiting the posts of added comments''} (S7, male, age: 22).

\subsection{Design Goals of \name{}} 
Based on the findings of the formative study and related work,
we derive the following design goals (DGs) of \name{}. 


\textbf{DG1: Sort posts based on the knowledge taxonomy about the UI components and visual elements that their comments mention.
} 
Existing ODCs usually provide a taxonomy that classifies posts based on their intentions or types of design examples, \eg flairs like ``UI/UX Design Feedback Request'' in Reddit \rr{\textit{r/UI\_design}} and filters like ``Mobile'' and ``Animation'' in Dribbble. 
However, few ODCs provide a knowledge taxonomy that could enable viewers to locate helpful information for learning their interested visual design concepts \cite{ely2000taxonomy,irvine2021taxonomies} (C1, C2). 
The different types of UI components and visual elements offer a good starting point for building such a knowledge taxonomy, which has been used in previous visual design tutoring tools \cite{DQ2023} or websites. 
\cxy{Our participants in the formative study also rated the features that sort posts based on mentioned UI components and visual elements as highly useful (\autoref{fig:formative_study} \rr{A1} \& \rr{A2} \& \rr{B1}). 
}

\textbf{DG2: Support easy navigation of the comments' content that offers constructive feedback and is related to learners' interested knowledge points.}
Navigating useful content in the comments is a common challenge in the exploration of online communities \cite{PlanHelper, Amplifying-the-music-listening-experience} and is also reported as challenging in our scenario (C3). 
Providing filters and highlighting key information in the comments are widely used in previous interactive tools for exploring online communities \cite{PlanHelper, Amplifying-the-music-listening-experience} and were rated potentially useful by our participants (\autoref{fig:formative_study} \rr{B2}). 
As suggested by \citet{DQ2023}, feedback that is constructive for learning purposes should be meaningful critique, suggestion, or rationale for the critique and suggestion regarding certain UI components and visual elements. 

\textbf{DG3: Support interactive knowledge note-taking on the explored comments.
}  
Taking high-quality notes, especially those with a clear logic flow, can help people better digest and manage newly learned information \cite{brandl2010nicebook,informal_notetaking_chi_2004, Kalnikaité_Whittaker_2008} (C4). 
Feedback from our participants also indicated that a note-taking panel would be useful for knowledge management (\autoref{fig:formative_study} \rr{B3}) and suggested that it should allow easy editing and review of the notes. 



%% file: sections/4_Design_and_Implementation.tex
\pzh{
\section{\name{} Interface Design and Implementation}
}

\pzh{
Based on the design goals derived from the formative study, we develop \name{}, a redesigned interface of ODCs for novices who browse ODCs to learn visual design.
\name{} sorts posts using a taxonomy of UI components and visual elements in the \textit{Overview} page and \textit{Recommendation} pane (DG1, \autoref{fig:User Interface} \xinyue{A and E}), filters and highlights the content in comments based on types of feedback and mentioned visual elements in the \textit{Comment} pane (DG2, \autoref{fig:User Interface} \xinyue{C}), supports interactive note-taking in the \textit{Note-Taking} pane (DG3, \autoref{fig:User Interface} \xinyue{D}).
We implement \name{} as a web app using the Vue framework to build the front end, a Python flask server to process user interaction, and a MongoDB database to store the structured posts and comments. 

Note that we do not intend \name{} to be a canonical interface for community members to share and discuss design-related content, but rather as a prototype to gain insights into the features that help novices learn by exploring content in ODCs. 
To keep our evaluation of \name{} focused, in the redesigned interface, we exclude other features like creating posts and comments, upvoting and downvoting, and rules of the community. 
In this section, we first present the knowledge taxonomy of visual design used in \name{}. 
Then, we describe a user scenario of \name{} and detail its interaction design and implementation. 
}

\peng{
\subsection{Taxonomy of Visual Design Knowledge}\label{sec:taxonomy}
}

\begin{table*}
    \centering
    \scalebox{0.77}{
    \begin{tabularx}{1.085\textwidth}{|c|c|c|c|c|c|c|c|}
        \hline
        \diagbox{UI Components}{Visual Elements} &\makecell{\textbf{Color}\\("white"\\"lighter")}&\makecell{\textbf{Contrast}\\("gradient"\\"transparent")}&\makecell{\textbf{Space}\\("padding"\\"gap")}&\makecell{\textbf{Typography}\\("font"\\"bold")}&\makecell{\textbf{Layout}\\("structure"\\ "alignment")}&\makecell{\textbf{Shape\&Size}\\("square"\\"width")}\\
        \hline
        \makecell{\textbf{Button}\\("home button", "settings button")}& 195& 174& 140& 79& 130& 135\\
        \hline
        \makecell{\textbf{Icon}\\("team logos", "website icon")}& 169& 109& 125& 72& 102& 116\\
        \hline
        \makecell{\textbf{Image}\\("photo", "pictures")}& 107& 104& 81& 45& 80& 86\\
        \hline
        \makecell{\textbf{Text}\\("paragraph", "labels")}& 166& 175& 135& 78& 128& 114\\
        \hline
        \makecell{\textbf{Background}\\("wallpaper", "color palette")}& 191& 172& 63& 54& 62& 64\\
        \hline
        \makecell{\textbf{Bar\&Page}\\("menu", "sidebar")}& 249& 209& 204& 107& 183& 184\\
        \hline
        \makecell{\textbf{\mr{Decorative Element}}\\("blocks", "sides")}& 166& 175& 136& 78& 128& 114\\
        \hline
        \makecell{\textbf{\mr{Interactive Card Element}}\\("Mastercard module", "portfolio")}& 72& 60& 51& 27& 61& 44\\
        \hline
    \end{tabularx}
    }
    \caption{
    The subcategories of UI components and visual elements in our taxonomy, \ie ``\textbf{UI components - a specific UI component} (\eg button) \textbf{- a specific visual element} (\eg color) \textbf{- a comment that this visual element and this UI component co-exist}''.  We detect and cluster keywords about UI design and calculate the number of comments that mention each type of UI component and visual element in our collected dataset. For example, if a comment contains the keywords ``gradient'', ``photo'', and ``home button'', we plus one to the number of comments that mention (Image, Contrast) and (Button, Contrast), respectively. 
    }
    \label{Taxonomy}
    
\end{table*}

\penguin{
We adopt two high-level categories, \ie UI components and visual elements, to build up our taxonomy of visual design knowledge in ODCs for two reasons. 
First, searching for design examples and knowledge about UI components in ODCs is a common practice \cite{li2014platform}, and commentary in ODCs commonly discusses the proper usage of visual elements within these components \cite{DQ2023,krishna2017increasing}.
Second, existing visual design taxonomies (\eg \cite{taxonamyui}, \cite{inbook}, Material Design \footnote{\url{https://m3.material.io/}}) widely include UI components and visual elements.  
For example, Material Design describes knowledge about color, shape, and typography under the ``Styles'' category, which is similar to our category of 
\xy{``visual element''}, and it talks about app bars, badges, and buttons under the ``Components'' category. 
Instead of directly leveraging the existing comprehensive taxonomies,
we choose a bottom-up approach and mine such sub-categories from the comments in ODCs as each ODC could mainly discuss certain sub-categories. 
For instance, compared to the buttons, the time pickers could rarely exist in the shared design examples in Reddit \rr{\textit{r/UI\_design}} community, and it would make it difficult to label and recognize the keywords that talk about time pickers in this ODC. 
}
\penguin{
We observe that the keywords about UI components and visual elements often co-exist in the knowledge points, \eg the color of a button should be red when it intends to mean ``stop''. 
To obtain a hierarchical taxonomy, we choose to prioritize UI components as visual elements could be more meaningful if they are contextualized by a specific UI component. 
Finally, the hierarchy of our intended taxonomy is ``\textbf{UI components - a specific UI component} (\eg button) \textbf{- a specific visual element} (\eg color) \textbf{- a comment that this visual element and this UI component co-exist}''. 
We acknowledge that our taxonomy built upon comments is less comprehensive compared to existing ones, and its comments may not explicitly reveal high-level design principles. 
However, we consider our taxonomy valuable as it can complement existing visual design taxonomies with rich peer-contributed feedback on the UI components and visual elements in numerous design examples. 
In the \autoref{sec:discussion_7.1}, we discuss possible ways to improve our taxonomy. 
}

\pzh{
We build up the taxonomy of visual design knowledge by applying the computational workflow developed by \citet{DQ2023} to our newly crawled dataset in Reddit \rr{\textit{r/UI\_design}} (\autoref{Reddit_introduce}). 
 Our crawled dataset contains 635 posts with the ``Feedback Request'' flair, with one image of a design example and with at least one comment (excluding the one from the bot AutoModerator and those from original posters), and there are 5523 comments under these posts. 
\penguin{After coding by one author and re-examining by another author, 312 (49.31\%) out of the 635 post images are about mobile app design, 205 (32.28\%) images are about web design, and the remaining  118 images are about other UI design tasks like \penguin{game interface, icon, and logo design}.}
\penguin{
We randomly sample 11 out of the 635 posts and annotate their associated 104 comments following the practices by \citet{DQ2023}. 
The two annotators discussed and resolved 11, 15, and 30 disagreement on the annotated sentences, UI component keywords, and visual element keywords. 
In total, we label 92 critique sentences, 135 suggestion sentences, 29 rationale sentences, 260 UI component keywords, and 121 visual element keywords. 
We use the fine-tuned RoBERTa-base model \cite{DQ2023} in the workflow to classify each of our labeled sentences into ``critique'', ``suggestion'' and ``rationale'', which achieves a 0.878 weighted F1 score. 
Meanwhile, we adopt the fine-tuned bert-base-cased model to recognize the keywords about UI components and visual elements in our labeled data, which achieves a 0.881 F1 score. 
}

\penguin{
We apply the validated models from \cite{DQ2023} to all 5523 comments, which yield 5398 critique sentences, 6223 suggestion sentences, 1378 rationale sentences, 7252 unique UI component keywords, and 4135 unique visual element keywords. 
}
Meanwhile, we adopt the fine-tuned bert-base-cased model, which achieves a 0.710 F1 score as reported in \cite{DQ2023}, to recognize the keywords about UI components and visual elements in comments. 
Next, we adopt ``all-MiniLM-L6-v2'', as used in the workflow in \cite{DQ2023}, to map all the recognized keywords into a 384-dimensional dense vector space and apply a K-means algorithm to cluster these keywords. 
We manually go through the output results with the $K \in [3, 10]$
and find that most of the clusters make sense with $K = 6$ for grouping visual elements and with $K = 8$ for grouping UI components. 
Members of our research team discuss the proper names for each cluster based on the prominent keywords in that cluster. 
\autoref{Taxonomy} summarizes the \penguin{named clusters} and the number of comments that mention each \penguin{sub-category} of UI component and visual element in our dataset.  
}

\peng{
\subsection{User Scenario}


\begin{figure*}
    \centering
    \includegraphics[width=1\linewidth]{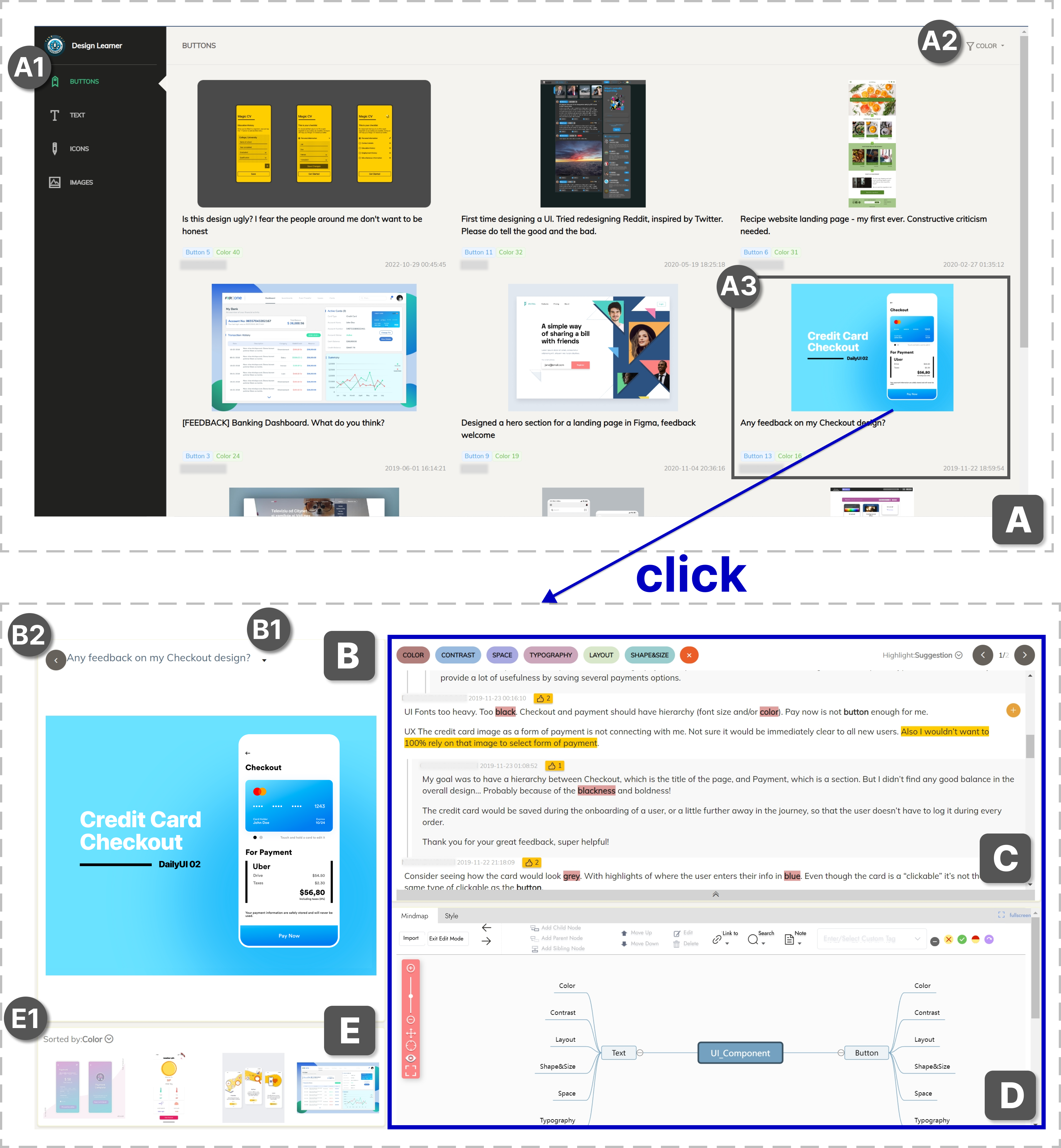}
    \caption{\xinyue{Interface of DesignLearner with a sample of UI design and mindmap. (A) Overview Page. (B) Image Pane. (C) Comment Pane. (D) Note-Taking Pane. (E) Recommendation Pane.}}
    \label{fig:User Interface}
\end{figure*}

\begin{figure*}
    \centering
    \includegraphics[width=1\linewidth]{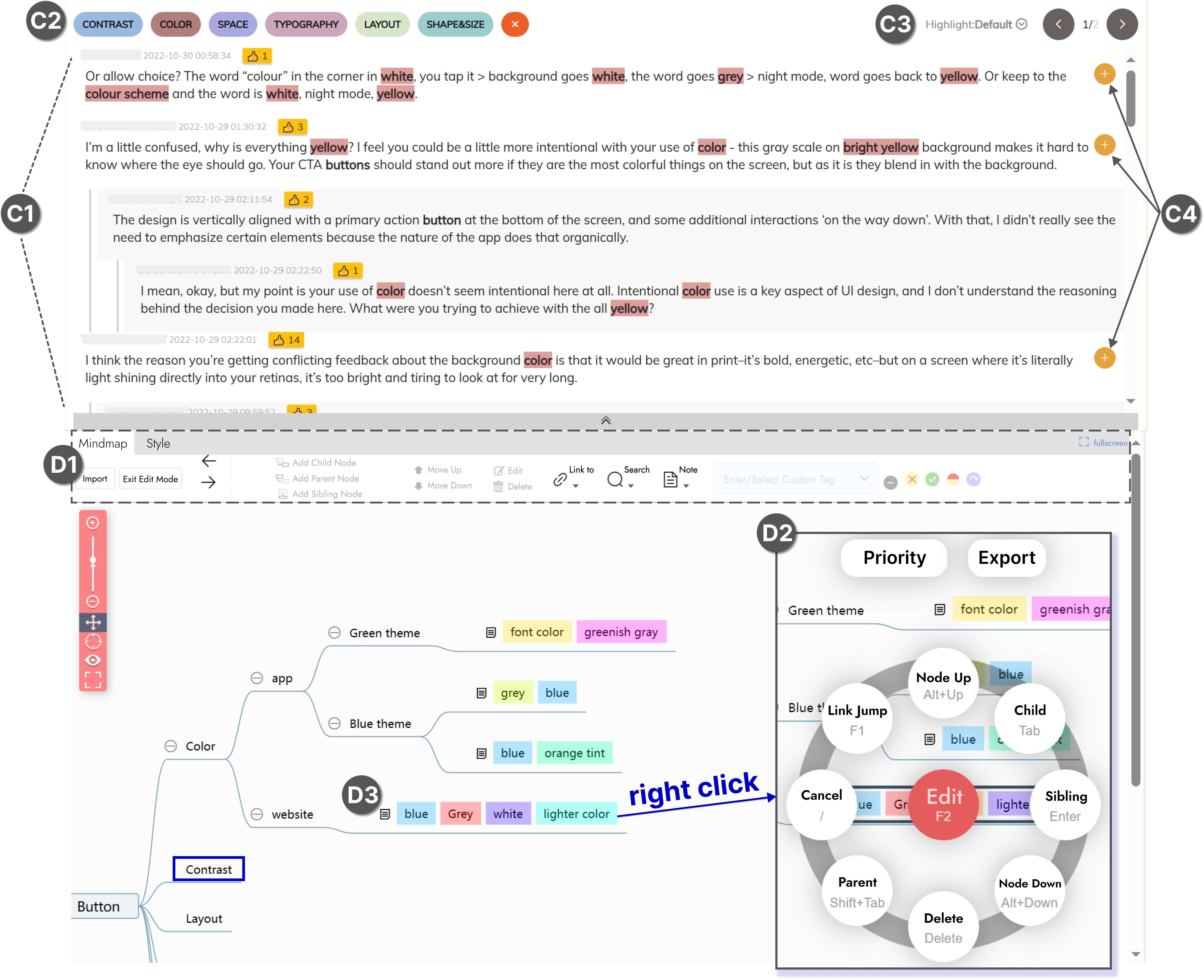}
    \caption{\xinyue{A detailed view of the Comment Pane and Note-Taking Pane in the second interface \rr{(the bottom of \autoref{fig:User Interface})}. (C1) Comments on the post. (C2) Filter for visual elements. (C3) Filter for feedback. (C4) Shortcut to take notes. (D1) Menu bar of the mind map. (D2) Pop-out menu for the right-clicked node. (D3) Notes associated with the node.}}
    \label{fig:Detail Page}
\end{figure*}

In this scenario, we describe how John, a male novice UI designer, learns about visual design knowledge about the color of ``button'' in \name{}. 
\peng{In the Overview Page of \name{} (\autoref{fig:User Interface} 
\xinyue{A}), John selects the UI component ``Button" \xinyue{(\autoref{fig:User Interface} A1)} and the visual element ``Color" \xinyue{(\autoref{fig:User Interface} A2)} to filter posts in the community. 
He browses the filtered posts, with a specific focus on the preview of the design example and the number of times that the post's comments mention ``Button'' and ``Color''.} 
\peng{
He clicks a post that he likes and proceeds to the reading page 
\xinyue{(\autoref{fig:User Interface} bottom).}
}
By looking at the highlighted keywords that mention the Color element, he quickly finds the comment that talks about how the color of the buttons is well-designed in the design example. 
He comes across another meaningful comment: \textit{``In addition to being consistent with the overall design tone, the color of the Button should be more prominent, catching the users' attention at first glance''}. 
He likes it and wants to take note of it, so he clicks the ``+'' button next to the comment, which adds the comment as a node to the mindmap (\autoref{fig:User Interface} \xinyue{D}). 
After examining the color-related comments in this post, John looks at the recommendation pane (\autoref{fig:User Interface} \xinyue{E}) and selects a post that also contains comments about the color of buttons in the design example. 
During the exploration of multiple posts, he continually records the helpful comments and takes notes in the mindmap. 
After the 30-minute learning session in \name{}, John feels satisfied and takes away a mindmap that documents what he learns from the community. 
}

\cx{
\subsection{User Interface and Implementation}
\pzh{
}

\pzh{
\subsubsection{Overview Page}
The overview page (\autoref{fig:User Interface} \xinyue{A}) serves as the home page similar to that of other ODCs (\eg Reddit \rr{\textit{r/UI\_design}} or Dribbble), which lists all posts in the community and enables users to filter the posts. 
Different from other ODC interfaces, \name{} supports users to filter posts based on the UI components and visual elements mentioned in the comments under the posts (\textbf{DG1}). 
Users can first select a UI component that they would like to learn knowledge about in the left menu bar (\xinyue{A1}). 
The selection of one UI component (\eg button) will sort the posts displayed in the overview page based on the number of the keywords (\eg home button, settings button) about that UI component mentioned in all comments under each post. 
Users can further filter posts by selecting their interested type of visual element (\eg color) in the drop-down menu located at the top-right corner of the page (\xinyue{A2}), which will resort the current posts based on the score calculated by weighting the number of targeted UI component and visual element keywords mentioned in the comments, \ie \peng{$score = 0.4 \times num\_UI\_component + 0.6 \times num\_visual\_element$}. 
\zhenhui{We choose these weights to emphasize the focus on the learning of visual elements attached to the given UI components in our later user study, while these weights could be re-assigned, \eg when \name{} serves learners with a focus on application of given visual elements to various UI components.}
For each sorted post, users can not only check its design example and title (\xinyue{A3}) as other ODCs support but also check the information about how many keywords in its comments talk about the targeted UI component and visual element. 
Users can click each post to check the design example and comments in detail on the following reading page.
}


\pzh{
\subsubsection{Reading Page}
The reading page (\autoref{fig:User Interface} \xinyue{bottom}) consists of four panes to help users digest the content of a post and learn visual design knowledge from it. 
The most unique differences between \name{} and other ODC interfaces are the comment pane that supports users to filter and highlight information based on visual element keywords (\textbf{DG2}) and the interactive note-taking pane for knowledge management (\textbf{DG3}). 
Besides, we customize the features in the image pane and recommendation pane that other ODC interfaces commonly have.  
}

\pzh{
\textbf{Image Pane} (\autoref{fig:User Interface} \xinyue{B}).
In this pane, users can click the image of the design example to view it in full screen, click the expanded icon to check the full-text body of the post title (\xinyue{B1}), and 
\cxy{get access to the original community by clicking on the title.}
\cxy{Besides, users can click the icon in the left or right of this pane to return to the previously explored post (\xinyue{B2}).}
}



\pzh{
\textbf{Comment Pane} (\autoref{fig:User Interface} \xinyue{C}). 
In this pane, users can filter and read the comments of their interests,  
\cxy{by scrolling through the comments pane or clicking the ``$\rightarrow$'' button (\xinyue{\autoref{fig:Detail Page} C1}).}
To filter comments and highlight the keywords related to a certain type of visual elements (\textbf{DG2}), users can click any button in the top menu bar of this pane (\xinyue{C2}). 
For example, if users click the brown ``Color'' button, \peng{the comments of the current post will be resorted based on the number of color-related keywords (\eg grey, white) each comment mentions, and the color-related keywords will be highlighted in brown in the comment. } 
Users can also select a feedback type (\ie critique, rationale, suggestion) in the drop-down menu \xinyue{(C3)}, which will highlight the related sentences in the comment to help them locate interested information. 
\cx{
At any time, users can click the ``x'' button to remove the highlights on keywords and select the ``default'' option in the drop-down menu to remove the highlights on sentences.
}
}

\pzh{
\textbf{Note-Taking Pane} (\autoref{fig:User Interface} \xinyue{D}). 
\label{Note-Taking-Pane}
In this pane, users can interactively add their learned knowledge as nodes of a mindmap and click each node to review the notes (\textbf{DG3}). 
\cx{We adapt vue-testcase-minder-editor \footnote{https://github.com/chenhengjie123/vue-testcase-minder-editor}. }
Users can manually add a node and type down their learned knowledge, thoughts, or related comments in the node at any level of the mindmap (\xinyue{\autoref{fig:Detail Page} D1}). 
\cxy{Alternatively, they can attach the comment to a selected node (\eg the Contrast node in a blue box in the mindmap of \autoref{fig:User Interface}) by clicking the ``+'' button adjacent to the comment (\xinyue{C4}). }
The name of the automatically added node is initialized by up to five randomly sampled keywords about UI components and visual elements in the ``+'' comment. 
Moreover, the automatically added node contains the ids of the post and the comment, and users can right-click the node and click the ``Jump'' button in the pop-up menu \xinyue{D2}), which will \cxy{refresh} the post and comments in the reading page based on the ids stored in the node.
Users can also hover over the note of the node to view the content of the associated comment and add any thought on the note whenever they want (\xinyue{D3}). 
The nodes in the mindmap will not change when users select another post and read it in detail. 
Lastly, users can export the constructed mindmap as a JSON, PNG, or JPG file, and import the JSON file to initialize the mindmap in the subsequent learning sessions (\xinyue{D2}).
}

\pzh{
\textbf{Recommendation Pane} (\autoref{fig:User Interface} \xinyue{E}).
In this pane, users can get recommendations for the next post to study, from which they could learn about the target visual design knowledge (\textbf{DG1}). 
With such a recommendation pane, users do not need to switch to the overview page to select another post, which could provide an immersive learning experience \cite{skulmowski2022understanding}. 
Specifically, the recommended posts are random samples of posts that have the same keywords for comments as the type of UI component users select on the overview page, and the same keywords as the type of visual element users select in the drop-down menu at the top of this pane (\xinyue{E1}). 
}
}

%% file: sections/5_Experiment.tex
\section{User Study}

We conducted a between-subjects user study with 24 novices of visual design to evaluate the effectiveness of user experience of \name{} for personalized visual design learning, compared with a baseline interface of Reddit \rr{\textit{r/UI\_design}}. 
\penguin{As described in \autoref{Reddit_introduce}, we chose \rr{\textit{r/UI\_design}}, because it provides rich design examples with UI components and constructive comments for our focused learning tasks.
In addition, participants in both groups can access similar types of design examples and comments as the data used in our demonstrated \name{} is sourced from \rr{\textit{r/UI\_design}}.}
The primary objectives of the user study are to assess whether and how browsing design examples and comments in \name{} affect learners' knowledge and skills of visual design. 
\penguin{Specifically, we focus on evaluating  \name{}'s features for personalized learning support, \ie taxonomy for adaptive content delivery and note-taking pane for interactive content management, which are unique compared to the baseline Reddit interface.}
Our research questions are:

\begin{enumerate}[label=\textbf{RQ}\arabic*]
  \item 
  How would \name{} affect the novices' learning outcome on \textbf{a)} knowledge of visual design and \textbf{b)} application of the knowledge in their visual design activity?
  
  \item How would \name{} affect the amount and the novices' perceptions of their explored design examples and comments in the learning process?
  
  \item How would \name{} affect novices' engagement and cognitive load in the learning session?

  \item How would novices interact with and perceive \name{} in the learning process?
  
\end{enumerate}

\subsection{Participants and Learning Tasks}
\label{sec:experiment-participants}
We recruited 24 participants (10 males, 14 females; age range 19-23, M = 20.38, SD = 0.95) from a local university via a post on social media. 
The inclusion criteria are that participants are novices in visual design but have an interest in learning more about it. We do not require the participants to be design students, because our assisted learning tool \name{} would support any students who are interested in learning visual design online. 
All of them are undergraduates and have passed the national College English Test for general requirements. Eleven of them major in Artificial Intelligence, five in Software Engineering, three in 
\xy{Microelectronics Science and Engineering}, two in 
\xy{Remote Sensing Science and Technology}
, one in Civil Engineering, one in 
\xy{Nuclear Engineering and Nuclear Technology}, and one in International Finance. 
In general, the majority of our participants are relatively inexperienced in learning UI visual design ($M = 2.375, SD = 1.81$) and exploring ODCs ($M = 2.958, SD = 1.88$; 1 - no experience at all, 7 - very experienced) with only three having worked on UI design projects and having experience in this area. But all of them are interested in learning it in our study (M = 5.417, SD = 1.1; 1 - not at all interested, 7 - very interested). The usage frequency of online communities
is: 13 daily, 6 2-6 days a week, and 5 once a week.

Participants were randomly allocated to either the \name{} (PD1-12) or Baseline group (PB1-12). 
Each participant was assigned the same learning task: 
``
As a visual design novice, you are going to learn about `button' and `text' UI components by exploring design examples and comments in an online design community to grasp the practical application of these concepts. 
Please use the given mindmap tool to record the content you find useful or the knowledge you learn, which will be used to assess your learning outcome.  
Before and after the learning session, you need to complete given design tasks that assess your knowledge about `button' and `text' in visual design.''  

\subsection{Measures}
\label{sec:measures}
{\bf RQ1. Learned knowledge of visual design and its application.} 
We measure participants' learned visual design knowledge in the ODC (\textbf{RQ1\_a}) by analyzing the mindmaps they created during the learning session. 
For each node in the mindmap, we assessed whether it revealed any meaningful knowledge point related to ``Button'' or ``Text'' (1 - yes, 0 - no), and if yes, we gave it a label about what concept it revealed, based on the literature (\eg \cite{jin2007touch, kim2017study, dodd2017designing}) and websites (\eg MaterialDesign) about UI design.
Three authors independently labeled a subset of nodes, discussed discrepancies, and established a coding scheme, which they applied to the remaining nodes,
without knowing whether they come from the \name{} or baseline condition. 
They reached a high level of agreement (Intraclass correlation coefficient = 0.9) and resolved the disagreement by majority vote. 
Finally, they worked together to determine and tag each node with labels of knowledge points, which turned out to be ``Contrast'', ``Color'', ``Typography'', ``Shape\&Size'', ``Layout'', ``Space'', ``Hierarchy'', ``Press Events (Button)'', and ``Literal Meaning (Text)''.


To assess \textbf{RQ1\_b}, we measured participants' performance in a visual design task with pre-test and post-test phases, following \cite{DQ2023}.  
In the \textbf{pre-test} before the learning task, participants were required to adjust the visual design of the \textbf{Button} and \textbf{Text} components in two given mockups (\autoref{fig:task1}a and \autoref{fig:task2}a). 
Designing mockups is a common graphical or visual design activity in design firms, and it requires designers to literally achieve the correct visual communication of user interface information using UI components (e.g., button, text) and visual elements (e.g., color, shape\&size, typography) \cite{Newman_Landay_2000}. 
\penguin{We followed two criteria to search mockups in the Figma design community. First, one mockup should be about mobile app design, and the other should be about web design, as these two are the most common designs in our \rr{\textit{r/UI\_design}} dataset (\autoref{sec:taxonomy}). Second, the mockups should feature buttons and text as the frequently appearing UI components, in line with the learning tasks.}
\penguin{We applied the first criterion to get five web and five mobile app mockups, and we followed the second criterion to select the two mockups \footnote{\url{https://www.figma.com/community/file/1074977530633823056} and \url{https://www.figma.com/community/file/ 1079825061633332655}} with the largest number of button and text components. }
We made the following adjustments to the mockups. 
First, we kept only one page in each of the original mockups. 
Second, we randomized the colors and fonts for all UI components except images. 
\penguin{Finally, we have one web page mockup \penguin{(four editable buttons, seven texts)} for a travel agency and one mobile payment page mockup \penguin{(four editable buttons, twelve texts)} for an educational application.} 

In the \textbf{post-test}, participants repeated the visual design task on identical mockups used in the pre-test to control for variations in design materials. Scores were assigned based on three core design principles from the Figma Library \cite{Principle}: \textbf{Hierarchy}, defined as "the mockup can help users recognize key information and distinguish it from the less important information";\textbf{ Consistency}, described as "the button and text applied in the UI are consistent and compatible with its brand"; and \textbf{Contrast}, characterized by "the button and text create a distinction between UI components". Ratings were made on a 7-point Likert scale, with 1 / 7 for strongly disagree / strongly agree. Two authors were familiar with the design principles\cite{Principle} and blinded to the test phase and group assignment, scored the anonymized and randomly ordered mockups. The final score for each aspect of the mockups was the average of their ratings.

{\bf RQ2 Explored design examples and comments.} 
With the users' consent, we recorded their computer screens in the learning session. 
With these records, we calculated the number of design examples and comments explored by each participant as follows: if the participant's web page remained in the unique example or comment for more than five seconds, we incremented the count by one. 
Besides, after the learning session, we measured participants' satisfaction and perceived helpfulness of the explored examples and comments using two items adapted from \cite{peng2020exploring, RQ1-2, DQ2023} on a 7-point Likert scale (1 / 7 - strongly disagree / strongly agree). \textbf{Satisfaction:} ``In general, I was satisfied with the design examples and comments I explored in order to learn about the buttons and text in UI design''. \textbf{Helpfulness:} ``Most of the examples and comments I explored were helpful for me to learn about the button and text in UI design''.

{\bf RQ3 Engagement and cognitive load in the learning process.} 
We adopted the metrics from \cite{DQ2023}, which were based on Brien's theoretical model \cite{Brien} and the flow theory for a positive experience \cite{Flow}, to assess user engagement in the learning process with six items (Cronbach's $\alpha =$ 0.772) on a 7-point Likert scale (1 / 7 - strongly disagree / strongly agree). 
They were \textbf{concentration} (``completely involved, focused, and concentrating''), \textbf{a sense of ecstasy} (``feel doing something is special''), \textbf{doability} (``skills are adequate, neither anxious nor bored''), \textbf{a sense of serenity} (``forgot about myself doing something''), \textbf{timeless feeling} (``time passed quickly''), and \textbf{intrinsic motivation} (``feel self-rewarded''). 
We measured users' \textbf{cognitive load} in the process using one 7-point Likert scale item ``I think the cognitive load of exploring design examples and comments to learn the visual design knowledge about button and text in this community is very low'' adapted from \cite{DQ2023}.

{\bf RQ4 Interaction and perception with \name{}.} 
To understand how participants used each feature of the redesigned interface in the learning process, we calculated the number of their clicks on each button in \name{}. 
As for the user perceptions with \name{}, we adapted the Technology Acceptance Model (TAM) \cite{TAM, TAM3} to measure the following in both the \name{} and baseline conditions: \textbf{usefulness} (four items, Cronbach's $\alpha = 0.833$); \textbf{easy to use} (four items, $\alpha = 0.686$); and \textbf{intention to use} (two items, $\alpha = 0.705$). 
The final score for each factor in the acceptance model was obtained by averaging the ratings across multiple items.

\subsection{Procedure}
\begin{figure*}[!ht]
    \centering
    \includegraphics[width=0.9\linewidth]{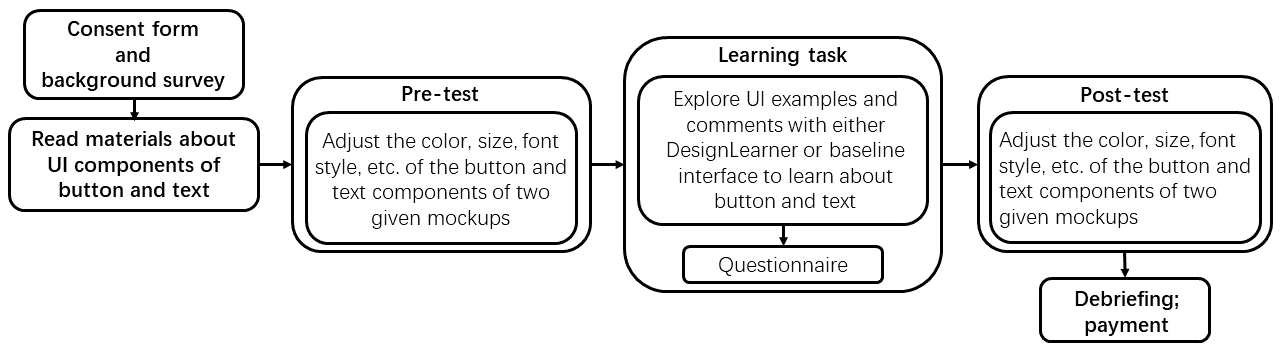}
    \caption{
    \rr{
    Procedure of the between-subjects experiment comparing the \name{} tool and Baseline Reddit interface.}
    }
    \label{procedure}
\end{figure*}

Figure \ref{procedure} illustrates the procedure of our user study. 
Participants used a desktop computer in our laboratory to conduct the study. Initially, they completed a consent form and a background questionnaire. To familiarize them with the fundamental concepts of buttons and text in document design, we provided instructional material compiled from references \cite{Button} and \cite{Text}. We then demonstrated how to use the Figma desktop application to adapt the UI components of the user interface mockup. Subsequently, participants engaged in a pre-test activity, which involved adapting two mockups from a visual design perspective. Based on a pilot study involving two users, we allocated 15 minutes for this \textbf{pre-test} phase. After the pre-test, we introduced the tools used for the learning task, i.e., \name{} or Baseline, and participants began exploring the online design examples and associated comments using the assigned tools for 25 minutes. Upon completion of the learning task, participants filled out a questionnaire asking them to 1) rate their perception of the learning process and the tool, and 2) comment on the given new UI design example in terms of buttons and text. Participants then performed a \textbf{post-test} of the visual design activity on the same two mockups as the pre-test in 15 minutes. Finally, we conducted brief interviews with participants, discussing their acquired knowledge, their perceptions of the tool's strengths and weaknesses, and their suggestions for improvement. The whole process lasted 60-90 minutes. At the end of the debriefing, each participant received a compensation of about $11$ USD.

%% file: sections/6_Analysis_and_Result.tex
\section{Analysis and Result}

To compare the numbers of learned knowledge points in RQ1\_a and the measured items in RQ2-4 between two user groups, we conducted Mann-Whitney U tests. 
To evaluate the changes in the participants' performance in the visual design activity (RQ1\_b), we conducted a two-way mixed ANOVA (between-subjects factor: \name{} vs. Baseline groups, within-subjects factor: pre-test vs. post-test). 
\autoref{Statistical_Results_of_RQ2-4} summarizes the statistical results of RQ1-4. 
As for the interview data, two authors transcribed it into text and conducted a thematic analysis. They first familiarized themselves by reviewing all the text scripts independently. After several rounds of coding with comparison and discussion, they finalized the codes of all the interview data regarding the pros and cons of \name{} and the baseline system (\autoref{Qualitative Results of RQ4}). 
\re{The quotes from the participants cited in the paper were translated from Chinese to English using Google Translate and then refined for grammar and completeness by two authors.}

\subsection{\textbf{Learning Outcome (RQ1)}}\label{knowledge points}
\subsubsection{Learned Knowledge Points (RQ1\_a)}
\begin{table*}
    \centering
    \scalebox{0.77}{
    \begin{tabularx}{1.295\textwidth}{c X X c c c}
        \hline
        & \name{} & Baseline & U & p & Sig.\\
        \hline
        Button & Color (24); Contrast (20); Layout (14); Typography (19); Shape\&Size (13); Space (13); Hierarchy (8); Events (10) & Color (17); Contrast (11); Layout (11); Typography (1); Shape\&Size (11); Space (8); Hierarchy (3); Events (0) & 11.50& 0.035 & \texttt{*}\\
        \hline
        Text & Color (11); Contrast (12); Layout (13); Typography (27); Shape\&Size (15); Space (9); Hierarchy (12); Literal Meaning (14)& Color (10); Contrast (9); Layout (6); Typography (9); Shape\&Size (9); Space (3); Hierarchy (6); Literal Meaning (4) & 2.50& 0.002 & \texttt{**}\\
        \hline
    \end{tabularx}
    }
    \caption{(RQ1\_a) Learned knowledge points about button and text with \name{} and the community-like baseline interface in experiment. The number next to each point is the number of comments in the mindmaps noted by participants; between-subjects; N = 24. We assess whether the knowledge points added to mindmaps contain meaningful knowledge concepts related to ``Button'' or ``Text''. If yes, we add a counter to the mentioned knowledge concepts. Note: \texttt{*}: \textit{p} < 0.05, \texttt{**}: \textit{p} < 0.01.}
    \label{Mindmap}
\end{table*}
\autoref{Mindmap} summarizes the comparative analysis of visual design knowledge acquisition between \name{} and baseline groups, revealing that
the average numbers of learned knowledge points were significantly larger in \name{} than in the baseline condition (Button: $U=11.50, p = 0.035$; Text: $U = 2.50, p = 0.002$), suggesting that \name{} can help users learn more knowledge points in the ODC.

Participants actively shared what they had learned from the ODC. 
For example, they learned that the buttons should be colored according to the actual situation. 
If designers want users to click on one call-to-action (CTA) button, they had better ``\textit{make the full-color option most prominent on places}'' (PD6, F, 21) or ``\textit{choose the color that will stand out more to users like yellow so they do what you want them to do}'' (PB3, M, 20).
Participants also learned to use button contrast to direct attention. 
``\textit{I would increase contrast — or even change the design of some things — to make things more visible}'' (PD10, M, 20). 
Besides, ``\textit{properly aligning and centering your elements is one of the most crucial steps in making your design look professional}'' (PD5, F, 20), which includes ``\textit{scaling down some of the less important elements like a few lines of text and add a little bit of spacing}'' (PD12, F, 20) and ``\textit{grouping similar actions into button menus while combining actions into a single action/button}'' (PB9, F, 21). 
They also learned some common practices in buttons. ``\textit{Red circles being the Record Button have been on VCRs, cassette recorders, and numerous other devices}'' (PB2, F, 21).

Regarding text, the knowledge points centered on font style and size
``\textit{The text normally is centered in the buttons}'' (PB7, F, 19). 
``\textit{Switching up the font weights on the icons and buttons can make them distinguishable}'' (PD1, F, 20). 
They also recorded some comments about 
how text style could be manipulated to attract attention and enhance readability,
``\textit{Off top of my head, try bold type, enlarge, and have it bleed off both sides slightly}''
(PD3, M, 21). 

\begin{figure*}[htbp]
    \centering
    \includegraphics[width=0.9\linewidth]{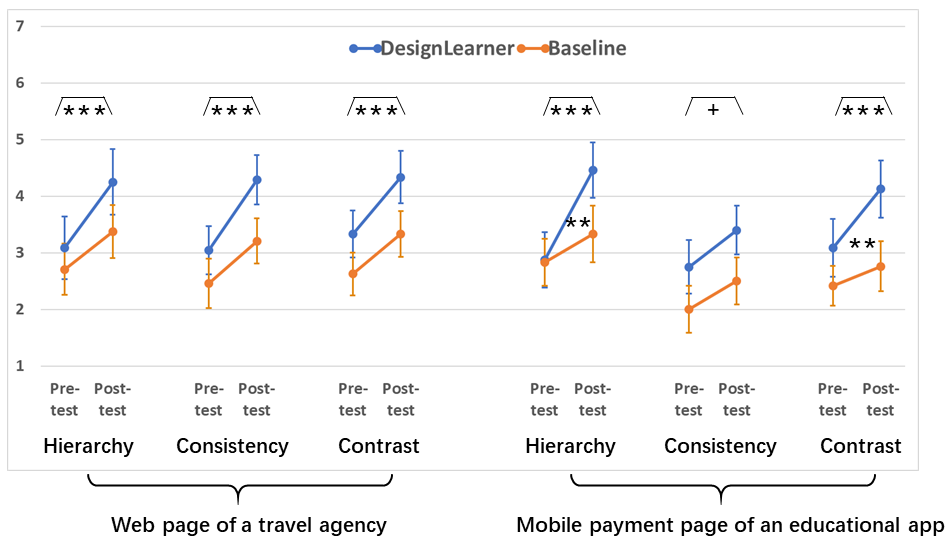}
    \caption{RQ1\_b results regarding the changes in participants' performance on the visual design activity before and after the learning session with either \name{} or Baseline interface. In both the pre-test and post-test, participants need to adjust the buttons and texts of two given mockups. Three authors rate the designs from aspects of hierarchy, consistency, and contrast, using a standard 7-point Likert scale, with 1 for strongly disagree and 7 for strongly agree. \texttt{***} : \textit{p} < 0.001, \texttt{*} : \textit{p} < 0.05, \texttt{+} : \textit{p} < 0.1.}
    \label{Figma}
\end{figure*}

\subsubsection{Application of Learned Knowledge (RQ1\_b)}
\begin{figure*}[h]
    \centering
    \includegraphics[width=1\linewidth]{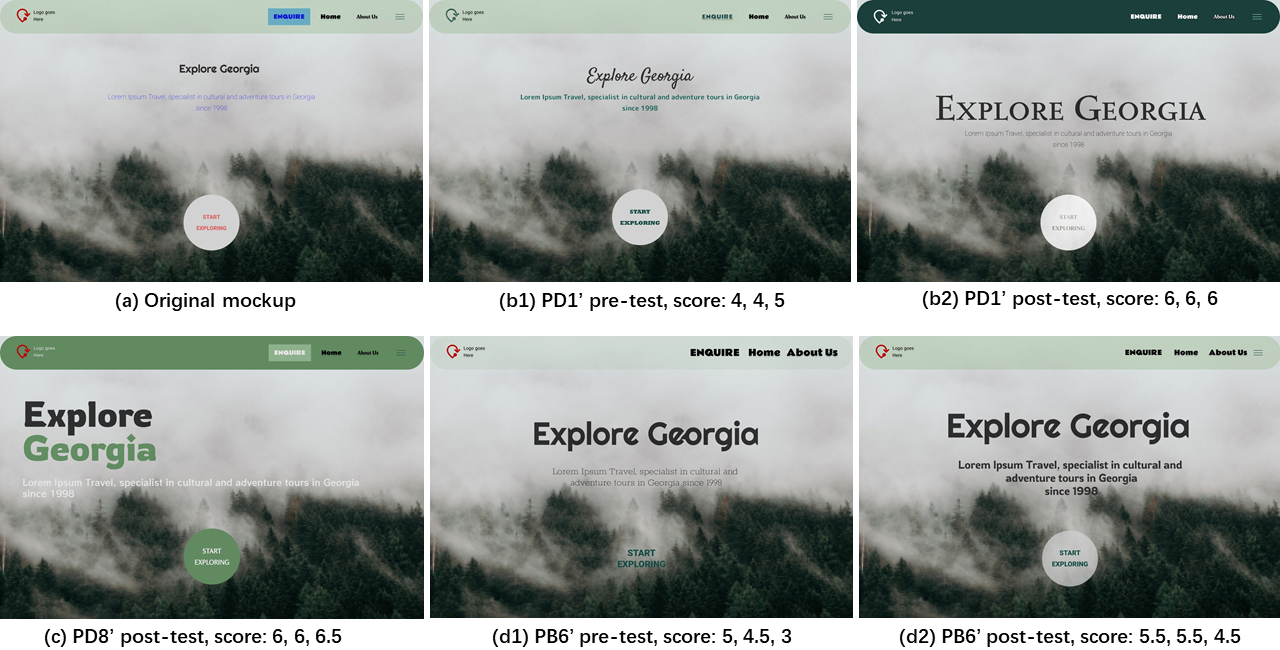}
    \caption{The original web page of a travel agency and the sampled outcomes from Participants in \name{} (note as PD) and Baseline (PB) groups. The average scores are from aspects of hierarchy, consistency, and contrast. This figure is better viewed in color.}
    \label{fig:task1}
\end{figure*}

\begin{figure*}[h]
    \centering
    \includegraphics[width=1\linewidth]{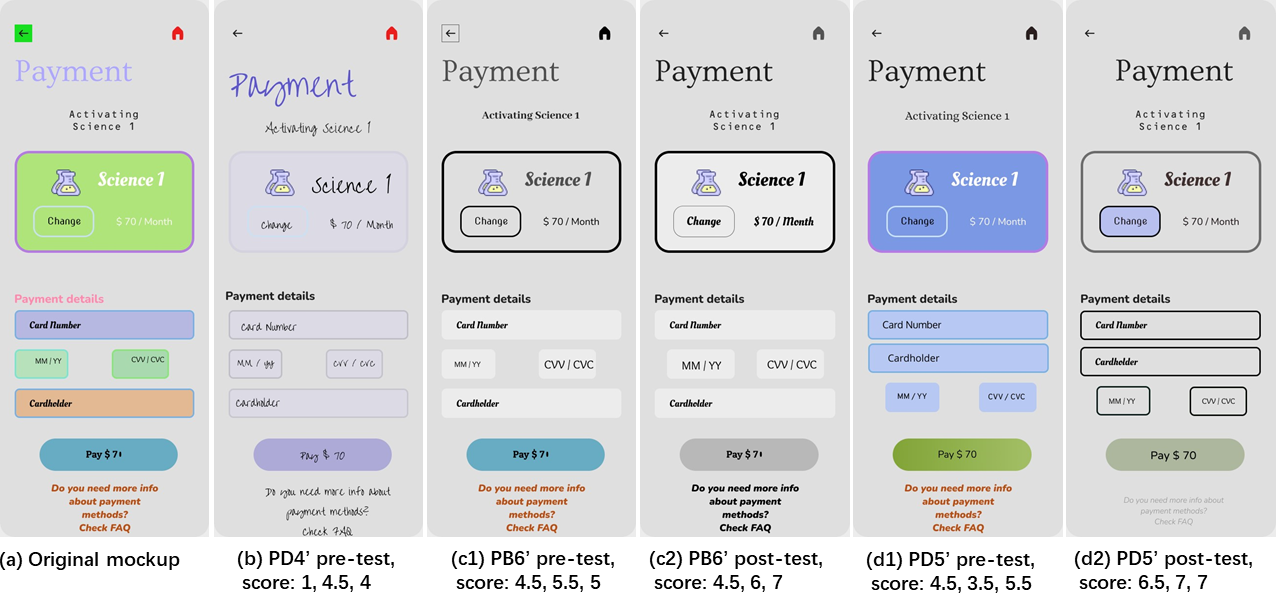}
    \caption{The original mobile payment page of an educational app and the sampled outcomes from Participants in \name{} (note as PD) and Baseline (PB) groups. The average scores are from aspects of hierarchy, consistency, and contrast. This figure is better viewed in color.}
    \label{fig:task2}
\end{figure*}

\autoref{Figma} demonstrates the enhancement in participants' visual design performance following a 25-minute learning task.
\re{The Mann-Whitney U tests indicate no significant differences between the two groups regarding participants' reported experience in UI design (p = 0.326) as well as their performance in pretests (Hierarchy: p = 0.341/0.817 for task 1/2, Consistency: p = 0.300/0.061, Contrast: p = 0.141/0.065). This confirms the success of our random assignment of participants.}
Regarding the visual design task on a travel agency web page (\autoref{fig:task1}) after learning with either \name{} or baseline interface, our findings reveal a significant improvement in participants' performances in matching the hierarchy (\textit{F} = 22.81, \textit{p} < 0.001), consistency (\textit{F} = 17.81, \textit{p} < 0.001), and contrast (\textit{F} = 22.08, \textit{p} < 0.001) design principles of button and text in the post-test over the pre-test. 
Neither the used interface nor its interaction with the time factor exerts a significant influence on the participants' performance in this visual design task. 
As for the visual design task on a mobile payment page (\autoref{fig:task2}), our results indicate a significant improvement regarding the measure of hierarchy (\textit{F} = 30.78, \textit{p} < 0.001), consistency (\textit{F} = 7.36, \textit{p} = 0.012) and contrast (\textit{F} = 26.26, \textit{p} < 0.001) in the post-test compared to the pre-test. 
The used learning interface does not significantly impact their performance on this mobile app page design task. 
However, we observe a significant interaction effect between the used interface and time factors on participants' performance in matching the hierarchy (\textit{F} = 14.87, \textit{p} = 0.003) and contrast (\textit{F} = 5.21, \textit{p} = 0.040) principles. 
This indicates that the degree of matching with the hierarchy and contrast principles among \name{} users improved significantly more than the baseline users after the learning session.


\autoref{fig:task2} shows the sampled outcomes of the visual design task on the mobile payment page. 
PB6 (M, 20) improved his performance after the learning session (\autoref{fig:task2} c1 and c2). 
One scorer judged PB6's pre-test design: ``\textit{The `PAY \$70' button and the red `more info' do not match the overall color style. Also, the font size is not handled well, key information such as `\$70/Month' does not stand out enough, and there are too many font styles in the input box}". 
PD5 (M, 20) also applied what he learned with \name{} to improve his design performance (\autoref{fig:task2} d1 and d2). One scorer commented on PD5's post-test design: ``\textit{The designer tried to create distinction and hierarchy with purpose. Yet, each component has a unique color and font type, making the page very inconsistent to read}". 

In summary, the results highlight the beneficial impact of ODC exploration on design performance, particularly noting a significant enhancement in mobile payment page design adherence to hierarchy and contrast principles in the \name{} condition over the Baseline. 
\re{\zhenhui{As detailed in the following subsections}, the performance enhancement observed in \name{}, as compared to the Baseline, can be attributed to two key factors. First, 
the availability of a richer set of UI design examples and comments in line with learning goals, accessible through filters (\autoref{6.2}).

Second, the interactive note-taking feature with comments facilitate easier documentation and retention of learned knowledge points from the community (\autoref{6.4.1}).}

\begin{table*}
  \centering
  \scalebox{0.77}{
  \begin{tabularx}{1.295\textwidth}{c X X}
        \hline
        \textbf{} & \textbf{\name{}} & \textbf{Baseline} \\
        \hline
        \textbf{Pros} & Mindmap (10); Keyword highlight (6); Filter post (5); Structured comments (5); Recommendation (5); Spilt screen (2); Sense of community (2) & Sense of community (8); Broad knowledge view (5); Clear and easy interaction (2) \\
        \hline
        \textbf{Cons} & Small comment pane (3); Complex interaction (2); Long automatically added notes (2) & Unstructured comments (8); Inaccurate search (6); Low-quality comments (5); Unstructured post (4); Lack professional comments (3) \\
        \hline
  \end{tabularx}
  }
  \caption{Summarized pros and cons of \name{} and the community-like baseline interface. The number next to each point is the number of participants who mention it; between-subjects; \textit{N} = 24.}
  \label{Qualitative Results of RQ4}
\end{table*}

\begin{table*}
    \centering
    \scalebox{0.77}{
    \begin{tabularx}{1.295\textwidth}{p{5.2cm} p{4cm} p{2.2cm} p{2cm} p{1.5cm} p{1.5cm} p{1.5cm}} 
        \hline
        \multirow{2}{*}{Research Question} & \multirow{2}{*}{Item} & \textbf{\name{}} & \textbf{Baseline} & \multicolumn{3}{c}{Statistics}\\ 
        \cline{5-7}
         &  &  Mean (SD)&  Mean (SD)&  U&  p& Sig.\\ 
         \hline
         \multirow{4}{=}{\parbox{5cm}{RQ2 Explored design examples and comments}} & Explored posts & 10.17 (4.45)& 4.00 (4.74) & 62.50 & 0.602 & \texttt{-} \\
         & Explored comments & \textbf{24.25 (8.85)}& 16.33 (6.82) & 35.00 & 0.034 & \texttt{*}\\
         \cline{2-7}
         & Satisfaction& 6.08 (1.00)& 5.33 (1.37)&  48.50&  0.174& \texttt{-}\\ 
         &  Helpfulness& \textbf{5.91 (1.00)}& 4.75 (1.36)&  36.50&  0.040& \texttt{*}\\ 
         \hline
         \multirow{8}{=}{\parbox{5cm}{RQ3 Engagement and cognitive load in the learning process}}&  Mean engagement& 5.76 (0.50)& 5.32 (1.18)&  50.00&  0.204& \texttt{-}\\ 
         &  - Concentration& 6.25 (0.97)& 6.00 (0.95)&  59.50&  0.470& \texttt{-}\\ 
         &  - Sense of Ecstasy& 6.33 (0.89)& 5.08 (2.15)&  51.00&  0.225& \texttt{-}\\ 
         &  - Doability& 5.67 (0.89)& 5.25 (1.22)&  57.50&  0.403& \texttt{-}\\ 
         &  - Sense of Serenity& 4.92 (1.56)& 4.25 (1.71)&  52.50&  0.260& \texttt{-}\\ 
         &  - Timelessness Feeling& 5.75 (1.14)& 6.25 (1.14)&  48.00&  0.166& \texttt{-}\\ 
         & - Intrinsic Motivation& 5.67 (0.78)& 5.08 (1.68)& 60.00& 0.488& \texttt{-}\\
         \cline{2-7}
         & Cognitive load& 5.91 (0.90)& 5.33 (1.23)& 53.50& 0.285& \texttt{-}\\ 
         \hline
         \multirow{3}{=}{\parbox{5cm}{RQ4 Interaction and perception with \name{}}}& Usefulness& \textbf{6.31 (0.49)}& 5.23 (1.00)& 24.50& 0.006& \texttt{**}\\ 
         & Easy to use& 5.56 (0.68)& 5.17 (1.10)& 59.00& 0.451& \texttt{-}\\ 
         & Intention to use& \textbf{6.42 (0.51)}& 5.33 (1.11)& 26.50& 0.008& \texttt{**}\\ 
         \hline
    \end{tabularx}
    }
    \caption{RQ2-4 statistical results about \name{} and the community-like baseline interface. All items except the number of posts and comments explored for RQ2 are measured using a standard 7-point Likert scale (1 - strongly disagree; 7 - strongly agree). Note: \texttt{-}: \textit{p} > 0.1, \texttt{+}: 0.05 < \textit{p} < 0.10, \texttt{*}: \textit{p} < 0.05, \texttt{**}: \textit{p} < 0.01; Mann-Whitney U test; between-subjects; \textit{N} = 24.}
    \label{Statistical_Results_of_RQ2-4}
\end{table*}

\subsection{\textbf{Explored Design Examples and Comments (RQ2)}}\label{6.2}
Through screen recordings, we determine that \name{} users explored more UI examples (Mean = 10.17, SD = 4.45 vs. M = 4.00, SD = 4.74)  and significantly more comments (M = 24.25, SD = 8.85 vs. M = 16.33, SD = 6.82; U = 35.00, p = 0.034) than those using the baseline tool during a 25-minute learning session, with higher perceived satisfaction and helpfulness scores (shown in \autoref{Statistical_Results_of_RQ2-4}).
These results indicate that in contrast to community-like interfaces, \name{} can effectively assist users in exploring helpful UI design examples and comments more efficiently for their personal learning goals.

Five participants appreciated the post-filters based on 
knowledge taxonomy, which assisted them in finding desired posts more efficiently. 
 ``\textit{The filters of visual elements (in the Overview page) helped me to find posts that contain comments about ``color'' in a short time}'' (PD10, M, 20). 
 Five participants also favored the comment filters based on visual components and types of feedback for helping them locate needed information. 
 ``\textit{The keywords highlighted in comments help me quickly locate relevant content in lengthy comments, which ensures I do not miss anything}'' (PD3, M, 21). 
 In contrast, in the baseline interface, two participants (PB5, PB9) felt uncomfortable when reading the comments, as they needed to frequently scroll up and down the page. 
 Another two participants in the baseline group interviews felt less attentive or even reluctant to read the comments.

\subsection{\textbf{Perception with Learning Process (RQ3)}}
\autoref{Statistical_Results_of_RQ2-4} reveals that
participants with \name{} report higher engagement levels (M = 5.76, SD = 0.50) than those using the baseline interface (M = 5.32, SD = 1.18; U = 50.00, p = 0.204), with particular improvements in feelings of ecstasy (M = 51.00, SD = 0.225) and serenity (M = 52.50, SD = 0.260) during the learning process. 
No significant differences are noted in other engagement aspects or cognitive load, suggesting that \name{} offers a more engaging experience without increasing cognitive workload for novices compared to community-like interfaces.

\subsection{\textbf{Interaction and Perception with \name{} (RQ4)}}

As shown in \autoref{Statistical_Results_of_RQ2-4}, participants using \name{} rate it significantly more useful (M = 6.31, SD = 0.51) for visual design learning than those using baseline interface (M = 5.33, SD = 1.11; U = 24.50, p = 0.006), and express a significantly higher intention to use it for future learning (M = 6.42, SD = 1.40) compared to the baseline users (M = 4.54, SD = 1.63; U = 26.50, p = 0.008). However, there is no significant difference in the ease of use between \name{} (M = 5.56, SD = 0.68) and the baseline interface (M = 5.17, SD = 1.10; U = 59, p = 0.451).


\subsubsection{Commonly used features in \name{}}\label{6.4.1}
Through screen recordings, we observe that \name{} users frequently utilize visual element filters in the comment pane (M = 14.33, SD = 4.76), and often click randomly to find comments with highlighted keywords.
Then they dig into the comments to digest the knowledge points and note them down. 
``\textit{The comments with a lot of highlighted keywords tend to represent the ones that contain the most knowledge and are worth learning.}'' (PD2, F, 20).
Besides, participants frequently click the ``+'' button (M = 8.17, SD = 4.00) under each comment to add it as a node to the mind map, as PD7(M, 23) stated, \textit{``Users can add a comment with selected keywords directly to the mindmap. The design is great." }
Moreover, users in \name{} often select posts in the recommendation pane (M = 5.25, SD = 4.97) to explore further design examples and comments, with PD1(F, 20) finding the feature \textit{``convenient to continually read relevant posts''}.

\subsubsection{Pros and cons} \label{sec:pros_cons}

Participants utilizing \name{} found the note-taking pane particularly beneficial for systematic learning, with 10 participants mentioning its clarity and utility for recording and reviewing, as PD3 (M, 21) described it as \textit{``great for long-term learning."} The `node jumping function' was also praised by PD8 (F, 20) for its convenience in quickly locating relevant advice.
Additionally, the UI component-filtered posts and structured comments are favored, with participants (N=5) each appreciating these features and participants (N=6) noting that highlighting knowledge keywords improves efficiency in finding meaningful information. 
PD11 (F, 20) highlighted the community atmosphere, stating it was \textit{``interesting to read and learn the meaningful comments of other members."}
However, some participants found the number of interactive features overwhelming, and the pane sizes are unreasonable. Three participants noting the comments and mindmap panes are too small and concave. Two participants also felt that the annotations that were automatically added to the mindmap were too wordy and they could not choose their own statements.

\subsubsection{Suggestions for improving \name{}} 
Participants provided several suggestions for enhancing \name{}, particularly for long-term use. They desired a customizable and draggable mindmap pane (\textit{N} = 5).``\textit{Just like when reading a paper, the page where you take notes doesn't `take up the screen’ for a long time, mindmap can be put away using widget hovering and only opened when you want to use them}'' (PD4, F, 20).  
In addition, they would like to see the comments that are automatically added to the mindmap can be selected as part of the statement (\textit{N} = 5), and they can be edited in the mindmap, \eg underlined, highlighted (\textit{N} = 3). 
One participant also proposed the inclusion of pictures in the notes. Furthermore, three participants suggested that \name{} should expand its filtering capabilities to include not only UI components and visual elements but also design scenarios, \eg the main interface of a website, mobile payment interface. 

%% file: sections/7_Discussion.tex
\section{Discussion}

\mr{
In this paper, we present \name{}, a redesigned interface for ODCs aimed at facilitating personalized visual design learning via adaptive content delivery and interactive content management. 
}
\name{} is our first attempt to offer a ``learning mode'' with a comment-based knowledge taxonomy and interactive note-taking to traditional online communities (\eg about programming \cite{informal_learning_chi2022} and health \cite{peng2020exploring,zhenhui2021}) whose user-generated content is suitable for learning purposes. 
Our between-subjects study highlights the strengths of our redesigned interface for facilitating users to explore UI examples and comments that are helpful to their learning goals more efficiently and help them better master the visual design knowledge, compared to a traditional ODC in Reddit. 
Participants attribute these benefits to \name{}'s filters of posts and comments based on a visual design knowledge taxonomy and interactive mindmap. 
These results provide empirical evidence that the structured comments are helpful for filtering needed information \cite{CoArgue} and that the active note-taking features are useful for digesting and managing the content in online communities \cite{kang2021metamap, PlanHelper}. 
Our work sheds light on how to structure the knowledge shared online and offer a learner-oriented interface for existing online communities. 

\subsection{Developing a Knowledge Taxonomy from User-Generated Content Online}
\label{sec:discussion_7.1}
\mr{Our \name{} adopts a taxonomy of visual design knowledge derived from comments within an ODC focused on UI design.}
\zhenhui{
In this taxonomy, many knowledge points, like those about typography, contrast, and color, are not only applicable to UI design but also extendable to various visual design tasks such as graphics, packaging, and poster design. 
These UI components and visual elements serve as a shared language across various fields of visual creation \cite{ambrose2011fundamentals}. 
\penguin{
However, other types of ODCs may have a different focus when members discuss their artworks, leading to a different taxonomy from our UI design. 
For instance, \rr{\textit{r/Design}}, which covers a broader spectrum of visual design topics, also talks about color and line, but a taxonomy for structuring the comments in this community may emphasize categories about branding strategies or
visual storytelling, which are relevant to poster or logo design.}}
To apply our knowledge taxonomy development methods to other ODCs or communities,
we would need 
a sentence classifier that identifies the type of constructive feedback, and a keyword detector and a cluster that identifies what the feedback is about. 
For example, in mental health communities like Reddit r/Anxiety, researchers could structure the user-generated content based on what types of social support users seek or provide \cite{peng2020exploring,zhenhui2021} and what topics they discuss. 

Our findings suggest that we could further enrich our visual design knowledge taxonomy. 
\penguin{
Our taxonomy does not explicitly specify the related design principles that the comments talk about.} 
Participants need to read filtered comments and take notes on good practices or considerations for the mentioned UI components and visual elements in the comments.
\penguin{We propose adding enacted design principles in our taxonomy, \eg ``UI components - a specific UI component - a specific visual element - a design principle related to this UI component and visual element - a comment that this visual element and this UI component co-exist''. }
As suggested by participants, these principles may include appropriate button colors for clickability and the layout of buttons and icons for a professional website. 
To build a model to extract the design principles, we may need to additionally label the relation between the mentioned UI components and visual elements in the comments. 
Alternatively, we could explore the capability and performance of the large language models to infer the revealed design principles from comments. 
\penguin{
Another promising way to enrich a knowledge taxonomy is generalizing the note-taking panel in \name{} to a social annotation tool. 
Future work could explore mechanisms that enable individual learners to share and discuss their notes with others, \eg attaching the notes to associated comments. 
Within such a social annotation tool, learners can learn from each other and collectively contribute design knowledge about UI components and visual elements, forming a knowledge graph for the ODC. 
}

\subsection{Design Consideration}

Based on our design practice and findings with \name{}, we derive two design considerations for designing a learner-oriented interface of existing online communities. 

\penguin{
\subsubsection{Support adaptive content delivery based on a knowledge graph mined from the communities} 
Our \name{} supports users to filter posts and comments that contain specific types of UI components or visual elements and recommends posts related to the current ones, which are commonly used features by our participants (\autoref{6.4.1}).  
They appreciated the post and comment filters based on knowledge taxonomy that assisted them in finding desired content efficiently (\autoref{6.2}). 
These results indicate that these adaptive content delivery features based on our taxonomy mined from the community accommodated participants' interests in the personalized learning task. 
Nevertheless, compared to the baseline interface, fewer participants reported a sense of community and broad knowledge view in \name{} condition (\autoref{sec:pros_cons}). 
This could be due to that \name{} emphasizes directing learners to specific content with filters but fails to provide them with overviews of the knowledge points in the community and in the current post thread. 
Therefore, we suggest future work to incorporate an interactive knowledge graph into the \name{} interface as another adaptive content delivery feature. 
This graph would be pre-built based on our taxonomy and support navigation of posts in the whole community and comments under a specific post. 
For example, the graph could work similarly to the Zoomable Posts panel in \cite{wu2024comviewerinteractivevisualtool} and enable users to view numbers of posts and comments that contain certain UI components and visual elements as well as navigate to these content. 
}

\subsubsection{Support interactive note-taking on any content of learners' interests}

\peng{
Our \name{} supports users to click a button to add any comment of their interests as a node, named by its mentioned UI components and visual elements, into the mindmap. 
\name{} also allows users to click a node to navigate to corresponding comments and posts. 
While most of our participants (9/12) appreciate that this note-taking feature helped them manage knowledge points, two participants (P5, P13) are concerned that the automatically added names and notes (\ie the content of the comments) of the node could be lengthy. 
We suggest that a more powerful note-taking pane in the learner-oriented community interface should support interactive note-taking on any selected content (\eg keywords, sentences, or even certain segments of the design example) of learners' interests. 
Besides, the note-taking pane should also be able to generate knowledge connections among the nodes.
For example, following the Concept Guide \cite{ConceptGuide}, the future interface can use a force-directed graph to generate a conceptual map for the content in ODC, where each node in the conceptual map represents a concept, and its radius represents the frequency of the term. 
}


\subsection{Limitation and Future Work}
Our work has several limitations that call for future work. 

\penguin{\subsubsection{Constructing taxonomy of visual design knowledge with a top-down approach and multimodal data}
First,} we adopt a bottom-up approach to build the knowledge taxonomy of visual design from the detected and clustered keywords in the comments. 
We observe that many detected keywords (\eg ``Sharp edges'') are not 
properly categorized into clusters.
Future work could refine the taxonomy via a top-down approach, \eg first defining the types of UI components and visual elements and then labeling comments and training classification models. 
Second, we build a knowledge taxonomy based on the comments. 
However, the design examples themselves, either in the form of images, videos, or audio, could also contribute to the knowledge taxonomy. 
Future work can investigate the incorporation of multimodal analyses in extracting knowledge points from design examples to enrich our comment-based taxonomy. 

\penguin{\subsubsection{Generalizing findings with diverse user groups and learning tasks.}
First,} our participants in the formative and user studies are university students who speak English as their second language and \zhenhui{major in engineering-related subjects}. 
\penguin{Second, all our participants were novices of visual design and not familiar with ODCs, and novices who regularly browse ODCs may perceive \name{} differently.}
More studies on diverse user groups could be applied to improve the generalization of findings on \name{}. 
\penguin{Third}, in the controlled user study, the personal learning goal is controlled to specific tasks, \ie learning about buttons and text. 
\penguin{Fourth, while we included both web and mobile app mockups in the user study and found that users can learn various visual design principles about buttons and texts with \name{} (\autoref{Mindmap}), we did not further examine if users could indirectly learn design principles across the web and mobile app modalities}. 
Future work should include a field study in which users can explore the redesigned interface with their own goals to \penguin{provide more in-depth findings on} \name{}'s user experience \peng{and effectiveness}.

\penguin{
\subsubsection{Complementing user-generated content with AI-generated content and professional resources}
\name{} leverages the structured comments, one type of user-generated content, in ODCs to support visual design learning. 
However, it could be frustrating if the comments themselves are difficult to understand. 
To step forward, we suggest that future learner-oriented interfaces should complement the user-generated content in online communities with additional learning resources. 
One promising approach is to leverage AI-generated content (AIGC), \eg produced by the recent large language models (LLMs), to help learners digest the comments. 
For example, inspired by \cite{jiang2023graphologue}, the interface can support users to select any content in the community and query the LLMs to explain, question, or elaborate on it. 
Moreover, recent advances in multi-modal LLMs, \eg GPT-4 with vision, have made it possible to generate an understanding of the UI design examples. 
However, we should also be aware of the potential risks, \eg factual accuracy, content quality, and biases, by introducing AI-generated content in the redesigned ODC interfaces.
Such risks could have a more negative impact on the learning process for novice designers compared to more experienced ones who are better equipped to evaluate its usefulness. 
To mitigate these potential risks, we could further include professional resources (\eg the documents on the Material Design website) that are validated by visual design experts in the redesigned interface. 
Future work could explore Retrieval-Augmented Generation techniques to combine UGC, AIGC, and professional resources into a learner-oriented ODC interface for personalized learning tasks. 
}




%% file: sections/8_Conclusion.tex
\peng{
\section{Conclusion}
In this paper, we propose \name{}, a redesigned interface of ODCs to facilitate personalized visual design learning tasks from shared posts and comments. 
We present a comment-based knowledge taxonomy to allow 
\name{} support users to filter posts based on a knowledge taxonomy constructed from comments and interactively take notes on the comments of their interests. 
We compare \name{} with a traditional ODC baseline interface in a between-subjects user study with 24 novices.
The results show that \name{} improves participants' efficiency and engagement in learning visual design from UI examples and comments in the community. 
Participants with \name{} can apply what they have learned to enhance their performance in the visual design activity. 
We offer insights for structuring the knowledge from the user-generated content shared online and provide design considerations for a learner-oriented interface of existing online communities. 
We hope our work will attract more researchers to leverage the resources in online communities to satisfy informal personalized learning needs. 
}

%% file: sections/0_Appendix.tex
\clearpage
\appendix